\documentclass[10pt]{article}
\usepackage[latin1]{inputenc}
\usepackage{amsmath}
\usepackage{amsfonts}
\usepackage{amssymb}
\usepackage{mathrsfs}

\usepackage{autobreak}
\allowdisplaybreaks

\usepackage{MnSymbol}

\usepackage[cal=boondox,scr=boondoxo]{mathalfa}
\usepackage{relsize}

\usepackage{float}
\usepackage{subfig}
\usepackage{fancybox,graphicx}
\usepackage{subfig}
\usepackage{caption}
\usepackage{color}
\usepackage{authblk}

\usepackage[colorlinks]{hyperref}

\usepackage{hyperref}

\newcommand{\doi}[1]{{doi:~\href{https://doi.org/#1}{\nolinkurl{#1}}}\rmFullStop}

\newcommand*{\rmFullStop}{\rmifnextchar{.}{}{}}

\makeatletter
\newcommand{\rmifnextchar}[3]{%
  \begingroup
  \ltx@LocToksA{\endgroup#2}%
  \ltx@LocToksB{\endgroup#3}%
  \ltx@ifnextchar{#1}{%
    \def\next{\the\ltx@LocToksA}%
    \afterassignment\next
    \let\scratch= %
  }{%
    \the\ltx@LocToksB
  }%
}
\makeatother 

\usepackage{accents}
\usepackage[titletoc,title]{appendix}
\usepackage{cite}

\usepackage{mathtools}

\DeclarePairedDelimiter\floor{\lfloor}{\rfloor}

\usepackage[top=2in, bottom=1.5in, left=1in, right=1in]{geometry}

\usepackage{smartdiagram}
\usesmartdiagramlibrary{additions}

\usepackage{tikz,xcolor}

\definecolor{lime}{HTML}{A6CE39}
\DeclareRobustCommand{\orcidicon}{%
	\begin{tikzpicture}
	\draw[lime, fill=lime] (0,0) 
	circle [radius=0.16] 
	node[white] {{\fontfamily{qag}\selectfont \scriptsize ID}};
	\draw[white, fill=white] (-0.0625,0.095) 
	circle [radius=0.007];
	\end{tikzpicture}
	\hspace{-2mm}
}

\foreach \x in {A, ..., Z}{%
	\expandafter\xdef\csname orcid\x\endcsname{\noexpand\href{https://orcid.org/\csname orcidauthor\x\endcsname}{\noexpand\orcidicon}}
}
\definecolor{myblue}{RGB}{9, 10, 95}
\definecolor{mygreen}{RGB}{9, 95, 20}

\usepackage{caption} 
\usepackage[cal=boondox,scr=boondoxo]{mathalfa} 
\usepackage{mathtools} 
\usepackage[titletoc,title]{appendix} 
\usepackage[left]{lineno}
\usepackage{setspace}

\title{Expansion Formula For the Magnetic Field of a Periodically Deformed Circular Current Loop\footnote{This is the version of the article before peer review or editing, as submitted by an author to \textbf{Expansion Formula For the Magnetic Field of a Periodically Deformed Circular Current Loop} IOP Publishing Ltd is not responsible for any errors or omissions in this version of the manuscript or any version derived from it. The Version of Record is available online at 
\url{https://doi.org/10.1088/1402-4896/ac1a4e} .}} 

\author[1]{Robert Salazar\orcidA{}}
\author[2]{Gabriel T\'ellez\orcidC{}}
\author[1,3]{Camilo Bayona-Roa\orcidB{}}
\affil[1]{Universidad ECCI, Bogot\'a, Colombia}
\affil[2]{Departamento de F\'isica, Universidad de los Andes - Bogot\'a, Colombia}
\affil[3]{Centro de Ingenier\'ia Avanzada Investigaci\'on y Desarrollo, CIAID - Bogot\'a, Colombia}

\begin{document}

    \maketitle

\begin{abstract}
A method is derived to obtain an expansion formula for the magnetic field  $\boldsymbol{B}(\boldsymbol{r})$ generated by a closed planar wire carrying a steady electric current. The parametric equation of the loop is $\mathcal{R}(\phi)=R+Hf(\phi)$, with $R$ the radius of the circle, $H \in [0,R)$ the radial deformation amplitude, and $f(\phi)\in[-1,1]$ a periodic function. The method is based on the replacement of the $1/|\boldsymbol{r} - \boldsymbol{r}'|^3$ factor by an infinite series in terms of Gegenbauer polynomials, as well as the use of the Taylor series. This approach makes it feasible to write $\boldsymbol{B}(\boldsymbol{r})$ as the circular loop magnetic field contribution plus a sum of powers of $H/R$. Analytic formulas for the magnetic field are obtained from truncated finite expansions outside the neighborhood of the wire. These showed to be computationally less expensive than numerical integration in regions where the dipole approximation is not enough to describe the field properly. Illustrative examples of the magnetic field due to circular wires deformed harmonically are developed in the article, obtaining exact expansion coefficients and accurate descriptions. Error estimates are calculated to identify the regions in $\mathbb{R}^3$ where the analytical expansions perform well. Finally, first-order deformation formulas for the magnetic field are studied for generic even deformation functions $f(\phi)$.
\end{abstract}

\section{Introduction}
In 1820, the French physicists Jean Baptiste Biot and Felix Savart derived an expression for the magnetic field due to an electrical current $i$ in a wire.
This expression has become fundamental in magnetostatics and it is now commonly known as the \textit{Biot-Savart law} \cite{griffiths2005introduction,jackson1999classical}.
It is
\begin{equation}
\boldsymbol{B}(\boldsymbol{r}) = \frac{\mu_o i}{4\pi} \oint_{\mathtt{C}} \frac{d\boldsymbol{r}' \times (\boldsymbol{r}-\boldsymbol{r}')}{|\boldsymbol{r}-\boldsymbol{r}'|^3},
\label{BCurveWireDefEq}    
\end{equation}
where $\boldsymbol{r}$ is \textcolor{black}{the observation point},  $\boldsymbol{r}'$ is the position of the wire differential segment, $\mu_o$ is the permitivity of the medium, and $\mathtt{C}$ is the path of the wire.
The depiction of this system is shown in Fig.~\ref{theSystemFig}.
The law is only valid for stationary currents, and it plays a similar role to that of Coulombs' law in electrostatics.

Although the Biot-Savart law is commonly used in magnetostatic, other physical problems require solving \textcolor{black}{integral expressions analogous to} Eq.~(\ref{BCurveWireDefEq}). One example is the case of vortex filaments in fluid mechanics \cite{van2015helical,kimura2018tent,ricca1995geometric,moin1986evolution,kondaurova2005full,adachi2010steady,gonzalez2019determination}, where the flow vorticity $\boldsymbol{\omega}=\nabla\times\boldsymbol{v}(\boldsymbol{r})$ is concentrated along a filament $C$, with $\boldsymbol{v}(\boldsymbol{r})$ the velocity field. 
For incompressible fluids satisfying $\nabla \cdot \boldsymbol{v}(\boldsymbol{r})=0$, the vorticity expression leads to the Poisson's problem $\nabla^2 \boldsymbol{v}(\boldsymbol{r}) = -\nabla\times\boldsymbol{\omega}(\boldsymbol{r})$\footnote{Being a direct consequence of $\nabla^2 \boldsymbol{v}(\boldsymbol{r}) = \nabla \left[\nabla \cdot \boldsymbol{v}(\boldsymbol{r}) \right] - \nabla\times\left[\nabla\times\boldsymbol{v}(\boldsymbol{r})\right]
$.}.

\begin{table}[h]
\begin{minipage}{.5\textwidth}
\begin{center}\small
    \begin{tabular}{ | p{3cm} | p{3.5cm} | }
    \hline
    \textbf{Vortex filaments} & \textbf{Magnetostatics} \\ \hline
    Velocity & Magnetic field \\
    $\boldsymbol{v}(\boldsymbol{r})$&$\boldsymbol{B}(\boldsymbol{r})/\mu_o$ \\ \hline
    Vorticity & Ampere's law\quad\quad\quad (Density current)\\
    $\boldsymbol{\omega}(\boldsymbol{r})=\nabla \times \boldsymbol{v}(\boldsymbol{r})$ 
    & $\boldsymbol{J}(\boldsymbol{r}) =\nabla \times (\boldsymbol{B}(\boldsymbol{r})/\mu_o)$\\ \hline
    Continuity & Gauss's Law \\
    $\nabla \cdot \boldsymbol{v}(\boldsymbol{r}) = 0 $
    &$\nabla \cdot \boldsymbol{B}(\boldsymbol{r}) = 0$ \\\hline    
    Circulation & Electric current \\
    $\Gamma = \int_S \boldsymbol{\omega}(\boldsymbol{r})\cdot d\boldsymbol{S}$ 
    & $i = \int_S \boldsymbol{J}(\boldsymbol{r})\cdot d\boldsymbol{S}$\\
    \hline
    \end{tabular}
    \caption {Physical analogy between magnetostatics and vortex filaments in incompressible fluid flows.}
    \label{vortexFilamentsManetostaticsTableEq}
\end{center}
\end{minipage}%
\begin{minipage}{.5\textwidth}
  \centering
\includegraphics[width=0.6\textwidth]{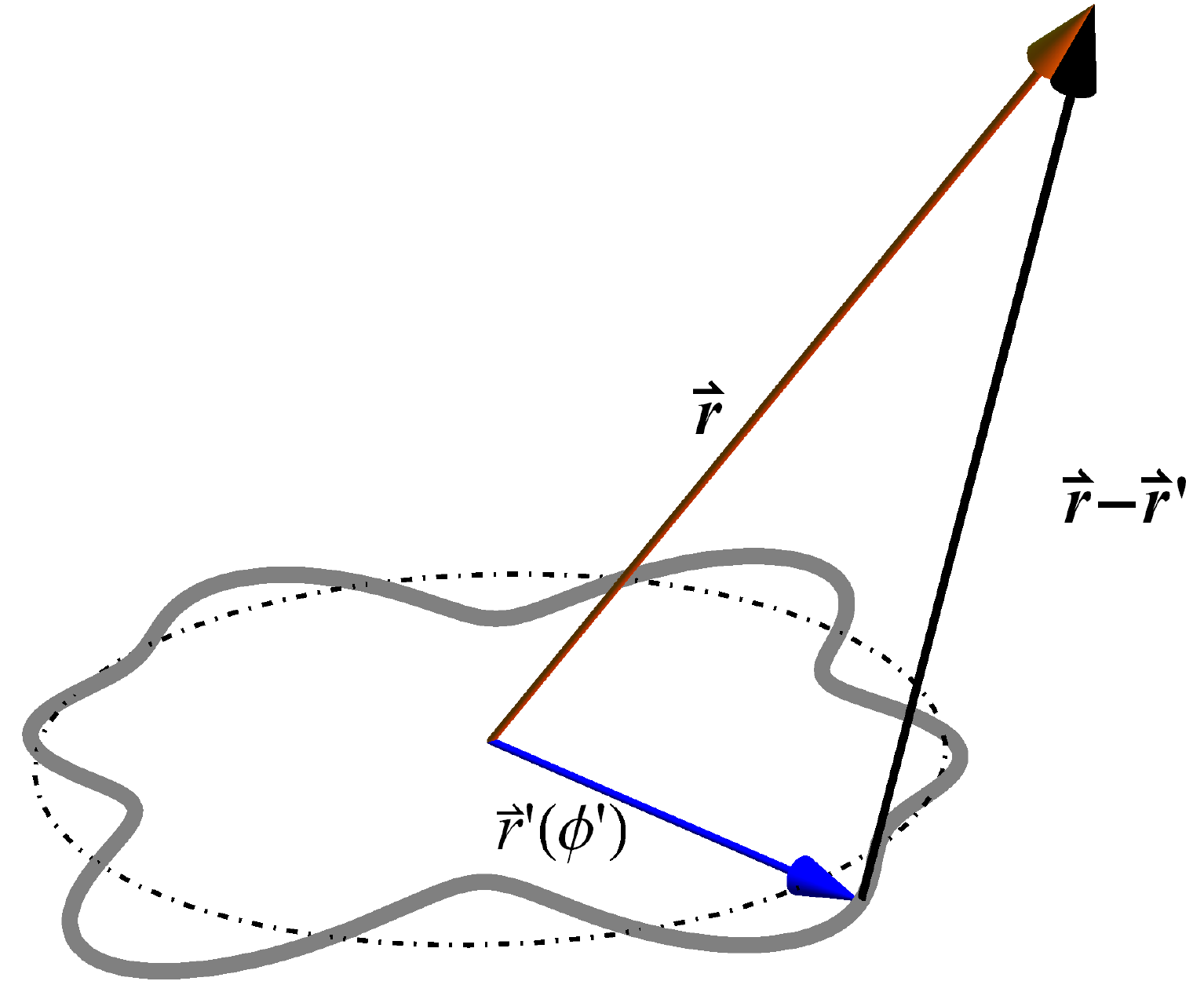}
\captionof{figure}{Generic loop.}
\label{theSystemFig}
\end{minipage}
\end{table}

Now, if the vorticity field is regular enough, \textcolor{black}{such that its derivatives are bounded and unique in $\mathbb{R}^3$,} then the solution of the previous Poisson's equation is the divergence-free velocity field
\[
\boldsymbol{v}(\boldsymbol{r}) = \frac{1}{4\pi} \int_{\mathbb{R}^3} \boldsymbol{K} (\boldsymbol{r}-\boldsymbol{r}') \times \boldsymbol{\omega}(\boldsymbol{r}') d\boldsymbol{r}'= \frac{\Gamma}{4\pi} \oint_{\mathtt{C}}  \frac{(\boldsymbol{r}-\boldsymbol{r}') \times d\boldsymbol{r}'}{|\boldsymbol{r}-\boldsymbol{r}'|^3}, \hspace{1.0cm}\mbox{with}\hspace{1.0cm}\boldsymbol{K} (\boldsymbol{r})=-\frac{\boldsymbol{r}}{4\pi |\boldsymbol{r}|^3},
\]
and $\Gamma$ the strength of the vortex. 
Table \ref{vortexFilamentsManetostaticsTableEq} lists the physical analogy between magnetostatics and the vortex filaments in fluid flows.

Even when the Biot-Savart law describes magnetostatic fields, it also plays and important role in the design and theory of wire antennas and radiation problems \cite{talashila2019determination,cintolesi2010modeling,li2015novel,galvis2016development}.  Another mathematically-equivalent physical system is the Gapless Surface Electrode (GSE). 
\textcolor{black}{That electrostatic system consists of an infinite conductor sheet laying on the $\mathbb{R}^2$-plane which has two regions that can be thought of as separated conductors (for example, metallic sheets which are very close to each other). The problem is to find the electrostatic field $\boldsymbol{E}(\boldsymbol{r})$ generated by two different potentials: one constant $V_o\neq0$ potential inside a closed region $\mathcal{A}$ and a grounded $V =0$ potential in the complementary $\mathbb{R}^2/\mathcal{A}$ region.}
The GSE problem can be solved from the evaluation of the following Biot-Savart-Like (BSL) expression:
\[
\boldsymbol{E}(\boldsymbol{r}) = \frac{V_o}{2\pi} \oint_{\partial \mathcal{A}} \frac{d\boldsymbol{r}' \times (\boldsymbol{r}-\boldsymbol{r}')}{|\boldsymbol{r}-\boldsymbol{r}'|^3},
\]
where $\partial \mathcal{A}$ is the border in between both electrodes. The previous expression becomes the electrostatic analog of the magnetostatic problem in (\ref{BCurveWireDefEq}).
The GSE serves as an ideal model of the gaped Surfaces Electrodes (SE), which are built technologies of Surface Electrodes\footnote{Indeed, the gaped SE solution is essential to describe SE Radio Frequency ion traps: a collection of consecutive SE that have become promising candidates for manufacturing large-scale quantum processors.} \cite{chiaverini2005surface,seidelin2006microfabricated,daniilidis2011fabrication,kim2011surface,hong2017experimental,mokhberi2017optimised,tao2018fabrication,tanaka2014design,zhang2020convenient,mount2013single}.

In the present work, analytic expressions are derived to represent the magnetic field generated by a closed planar electrical wire carrying a steady current.
Particularly, the three-dimensional vector field due to a circular-deformed loop with low symmetry.
Analytic solutions of the BSL integrals are outstanding for solving \textcolor{black}{steady electric, magnetic or fluid flow problems,} as it has been demonstrated in \cite{salazar2019AngularDependentSE, salazar2020gaped} by the authors of the present study \footnote{For instance, the recent methodology in \cite{salazar2020gaped} demonstrates how the electric field of Gaped Surface Electrode systems can be obtained from the weighted average of their gapless counterparts. In that work, it was formally demonstrated that the electric field of the Gapped surface electrode with gap $\mathscr{G}$ is given by
\[
\boldsymbol{E}(\boldsymbol{r}) = \frac{V_o}{2\pi} \int_{\mathscr{G}}  \frac{\boldsymbol{\mathscr{W}}_{\nu}(\boldsymbol{r}')d^2 \boldsymbol{r}' \times (\boldsymbol{r}-\boldsymbol{r}')}{|\boldsymbol{r}-\boldsymbol{r}'|^3} 
\]
and it is analogous to the magnetic field due to a ribbon displayed on $\mathscr{G}$ carrying a current surface density $\boldsymbol{K}$
\[
\boldsymbol{B}(\boldsymbol{r}) = \frac{\mu_o}{4\pi} \int_{\mathscr{G}}  \frac{\boldsymbol{K}(\boldsymbol{r}')d^2 \boldsymbol{r}' \times (\boldsymbol{r}-\boldsymbol{r}')}{|\boldsymbol{r}-\boldsymbol{r}'|^3},
\] 
where the weight vector $\boldsymbol{\mathscr{W}}_{\nu}(\boldsymbol{r}')$ depending on the gradient of the electric potential on the $\mathscr{G}$ is the electric counterpart of $\boldsymbol{K}$. Additionally, both fields satisfy analogous continuity conditions $\nabla \cdot \boldsymbol{K}(\boldsymbol{r}) = 0$ for charge conservation, and $\nabla \cdot \boldsymbol{W}(\boldsymbol{r}) = 0$.
}.

\textcolor{black}{
In addition to the Legendre polynomial expansion for the magnetic field of a current in a circular loop, the problem is exactly solvable in the case of the ellipse geometry of the wire \cite{urankar1985vector}.
However, there is no exact analytical solution for a generic curved loop. 
Direct numerical integration of the Biot-Savart formula over $\mathtt{C}$ is one possible solution procedure.
Correspondingly, numerical and expansion solutions for the perfect circular loop have been obtained in \cite{papakanellos2007alternative,fikioris2008use,papakanellos2009efficient}. Also, for the electrostatic analog problem of Surface Electrodes, numerical methods have been applied in \cite{salazar2019AngularDependentSE}. 
}


The approach in the present work makes use of the Gegenbauer Polynomials as an expansion basis for the three-dimensional magnetic field. 
This method stands out for the general deformation of the wire, given any deformation function $f(\phi)$ vanishing the symmetry.
Hence, the Legendre polynomial expansions of the vector potential $\boldsymbol{A}$ becomes less useful than the present approach since all the three-dimensional components of the vector potential become active and contribute\footnote{For symmetric problems, it could be convenient to find expansions of the vector potential $\boldsymbol{A}$ employing Legendre polynomials since there are components of $\boldsymbol{A}$ that reduce to zero. This is especially the case of the magnetic field for the circular loop.}\footnote{Additionally, the intermediate step of computing the magnetic field from the rotational operator of the vector potential is not necessary when adopting the Gegenbauer Polynomials.}.

The particular case of the $\mathcal{R}(\phi)=R + Hf(\phi)$ loop eventually allows to split the solution into two terms: the solution of the symmetric circular loop and the contribution of the deformation. 
This is advantageous because the magnetic field of the circular loop can be known exactly from classic results.
With this in hand, the method in this document focuses on determining the contribution of the magnetic field from the deformed part. 
As will be disclosed later, the expansion formulas can be reduced to compute integrals involving the generic $f(\phi)$ function. 
Moreover, this contribution can be easily handled in some ideal scenarios such as the harmonic deformation, where all the expansion terms can also be found exactly.


The remaining parts of this manuscript are organized as follows.
In Section~\ref{TheProblemSection}, the mathematical setting of the circular-deformed wire problem is recalled.
Next, the analytic methodology using the expansion approach with the Gegenbauer Polynomials is presented in Section~\ref{expansionSection}, which ends up giving the magnetic field expression.
Illustrative examples of the magnetic field calculation around circular-deformed wires are presented in Section~\ref{sec:examples}.
Next, the first-order description for a generic even deformation function is explained in Section~\ref{sec:GenericDeformation}. Finally, some conclusions are stated in Section~\ref{sec:conclusions}.

\section{The problem}
\label{TheProblemSection}
In the present section, the problem set of the circular-deformed wire $\mathcal{R}$ carrying a constant current $i$ is explained.
Certainly, the steady magnetic field given by Eq.~(\ref{BCurveWireDefEq}) can be written in terms of the spherical coordinates system $(r,\theta,\phi)$: being $r$ the radial distance, $\theta$ the azimuthal angle, and $\phi$ the polar angle, as follows,
\[
B_r(\boldsymbol{r}) = \frac{\mu_o i}{4\pi} \cos\theta \int_0^{2\pi} \frac{\mathcal{R}^2(\phi')}{|\boldsymbol{r}-\boldsymbol{r}'|^3} d\phi', \hspace{0.2cm} B_\phi(\boldsymbol{r}) = -\frac{\mu_o i}{4\pi} r\cos\theta  \int_0^{2\pi} \frac{ \dot{\mathcal{R}}(\phi')\cos(\phi-\phi') + \mathcal{R}(\phi')\sin(\phi-\phi') }{|\boldsymbol{r}-\boldsymbol{r}'|^3} d\phi',
\]
and
\[
B_\theta(\boldsymbol{r}) = \frac{\mu_o i}{4\pi} \left[r \int_0^{2\pi} \frac{\mathcal{R}(\phi')\cos(\phi-\phi')}{|\boldsymbol{r}-\boldsymbol{r}'|^3} d\phi' - \sin\theta\int_0^{2\pi} \frac{\mathcal{R}^2(\phi')}{|\boldsymbol{r}-\boldsymbol{r}'|^3} d\phi' - r \int_0^{2\pi} \frac{\dot{\mathcal{R}}(\phi')\sin(\phi-\phi')}{|\boldsymbol{r}-\boldsymbol{r}'|^3} d\phi'\right],
\]
where 
\begin{equation}
|\boldsymbol{r}-\boldsymbol{r}'| = \sqrt{r^2 + \mathcal{R}^2(\phi') - 2r\mathcal{R}(\phi')\sin\theta\cos(\phi-\phi')}.    
\label{rMinusrpEq}
\end{equation}
The magnetic field for the simple case of a circular loop $\mathcal{R}=R$ can be written in terms of the complete elliptic integrals of the first and second kind, $K$, and $\underbar{E}$, respectively. 
Hence, the magnetic field in terms of the spherical components for the circular loop is given by 
\begin{align}
    (B_r(\boldsymbol{r}))_{circle} &= \frac{R^2 i}{4\pi} \frac{4\cos\theta}{\mathscr{r}_{-}(r,\theta)^2 \mathscr{r}_{+}(r,\theta)}  \underbar{E}\left[ \frac{4rR\sin\theta}{ \mathscr{r}_{+}(r,\theta)^2 }\right], \label{magneticcircle1}\\
    (B_{\theta}(\boldsymbol{r}))_{circle} & = 2\frac{i}{4\pi} \frac{\csc\theta}{\mathscr{r}_{-}^2 \mathscr{r}_{+}} \left[  (r^2+R^2\cos(2\theta))\underbar{E}\left( \frac{4rR\sin\theta}{ \mathscr{r}_{+}^2 }\right) - \mathscr{r}_{-}^2 K\left( \frac{4rR\sin\theta}{ \mathscr{r}_{+}^2 }\right)\right]  \label{magneticcircle2},\\
    (B_{\phi}(\boldsymbol{r}))_{circle} & = 0,\label{magneticcircle3}
\end{align}

where $\mathscr{r}_{\pm}(r,\theta)$ is defined as $\mathscr{r}_{\pm}(r,\theta) = \sqrt{r^2 + R^2 \pm 2rR\sin\theta}$. \textcolor{black}{It should be noted that the $\phi$-component of the magnetic field vanishes because the system is axially symmetric.} Nevertheless, the general case $\mathcal{R}(\phi) = R + Hf(\phi)$ of a deformed circular wire $\mathcal{R}(\phi)$ with base radius $R$ and deformation amplitude $\nu=H/R \in (-1,1)$ is more ambitious.
Certainly, the magnetic field in the deformed-circular wire is a challenging problem that will be addressed next in the remaining sections of this article.
The key to doing this is that the magnetic field can be constructed from $\boldsymbol{B}(\boldsymbol{r}) = (\boldsymbol{B}(\boldsymbol{r}))_{circle} + \boldsymbol{B}^{(\nu)}(\boldsymbol{r})$, with $\boldsymbol{B}^{(\nu)}(\boldsymbol{r})$ the contribution of the deformation.

\section{Expansion approach}
\label{expansionSection}
In this section, the expansion solution of the magnetic field generated by a circular-deformed wire is achieved.
First, the expansion expression for the circular-deformed loop magnetic field is explained.
In this regard, the inverse distance term related to Eq.~(\ref{rMinusrpEq}) is first addressed and then used to find the magnetic field expression of the circular-deformed wire. Since the Gegenbauer Polynomials will be used as the basis of the expansion, they are summarized in Appendix Section \ref{GegenbaurePolyAppendixSection}. Next, we use an expansion solution in a particular case of harmonic deformations. 
An analysis of the first-order contribution of the deformed wire due to functions with even parity is presented at the end.

\subsection{Inverse distance}

\begin{figure}[h]
\begin{minipage}{1.0\textwidth}
  \centering 
  \includegraphics[width=0.4\linewidth]{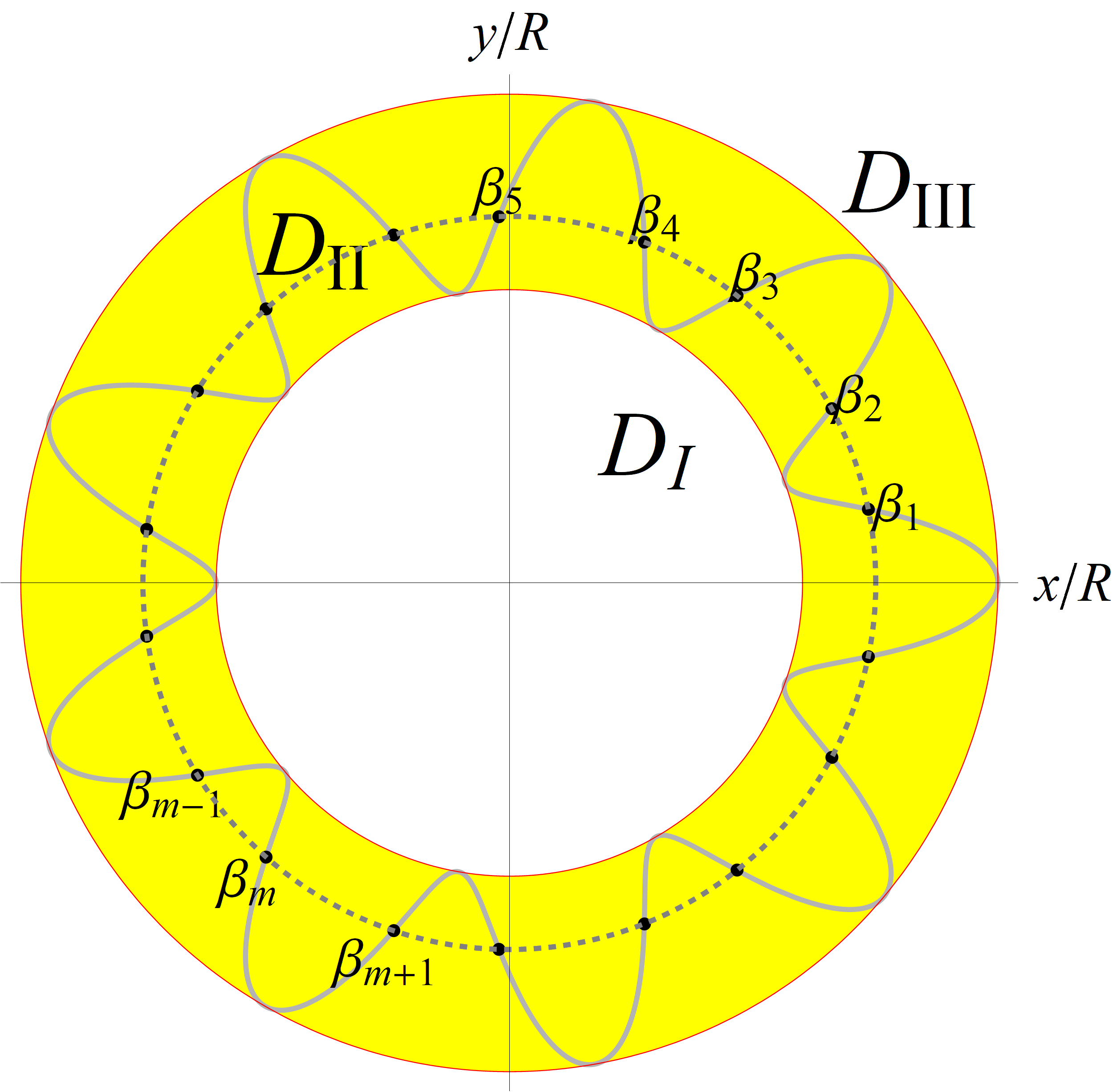}
\end{minipage}
\captionof{figure}{Regions.}
\label{regionsFig}
\end{figure}

Yet, the inverse distance in the circular-deformed wire problem can be calculated from the Gegenbauer polynomials.
To do so, the variable $\xi(\theta,\phi-\phi')=\sin\theta\cos(\phi-\phi')$ can be defined in terms of the angular coordinates.
Assuming that $1/|\boldsymbol{r}-\boldsymbol{r}'|^\alpha$, with $\alpha\in\mathbb{N}^{0}$, is evaluated in the region of the space where $r>\mathcal{R}(\phi')$, this inverse distance can be written in the more convenient way:
\[
\frac{1}{|\boldsymbol{r}-\boldsymbol{r}'|^\alpha} = \frac{1}{r^{\alpha}} \frac{1}{(1+2\chi(r,\phi')\xi(\theta,\phi-\phi')+\chi(r,\phi')^2)^{\alpha/2}} = \sum_{n=0}^{\infty} C_{n}^{\alpha/2} (\xi) \chi^n,
\]
 by introducing the Gegenbauer polynomials.
Unfortunately, there are regions in $D = \mathbb{R}^3$ where the previous expansion does not converge.  In order to deal with this problem, the region $D$ can be break up in three non-overlapping regions, as it is shown in Fig.~\ref{regionsFig}. 
\color{black}

The inner spherical region $D_I = \left\{(r,\phi',\theta) | 0 < r < r_{\min} \wedge 0 \leq \phi' \leq 2\pi \wedge 0 \leq \theta \leq \pi \right\}$, the outer region $D_{III} = \left\{(r,\phi') |  r \in [r_{\max},\infty) \wedge 0 \leq \phi' \leq 2\pi \wedge 0 \leq \theta \leq \pi \right\}$, and the intermediate region $D_{II} = D / D_I\cup D_{III}$, where $r_{\min}$ and $r_{\max}$ are the global extreme values of $\mathcal{R}(\phi')$. 
\color{black}
The intersection of the hollow sphere region $D_{II}$ with the $xy$-plane is yellow-highlighted in Fig.~\ref{regionsFig}.

Let the loop in the region is described by the roots $\Omega(r) = \left\{\beta_1,\beta_2,\ldots,\beta_M\right\}$ of $r-\mathcal{R}(\phi') = 0$, where $M$ is the total number of roots in the set $\Omega(r)$.
The condition $\beta_{M+1}:=\beta_{1}$ applies, as well.
Thus, let $\Xi_m = \left[\beta_m,\beta_{m+1}\right]$ be the angular intervals defined by the roots $\Omega(R)$ in the loop.

The function $\chi(r,\phi')$ can be defined inside $[0,\infty)\times[0,2\pi] \ni (r,\phi')$ as follows:

\[
\chi(r,\phi') := \frac{\mbox{min}(\mathcal{R}(\phi'),r)}{\mbox{max}(\mathcal{R}(\phi'),r)} =
      \frac{\mathcal{R}(\phi')}{r}\hspace{0.25cm}\mbox{\textbf{if}}\hspace{0.25cm}\vartheta(r,\phi')\hspace{0.25cm}\mbox{\textbf{otherwise}}\hspace{0.25cm}\frac{r}{\mathcal{R}(\phi')} ,
\]
where $\vartheta(r,\phi')$ is a condition defined as
\[
\vartheta(r,\phi') := \begin{cases}
       \mbox{True} &\quad\mbox{\textbf{if}}\hspace{0.25cm}\mathcal{R}(\phi')<r,\\
        \mbox{False} &\quad\mbox{\textbf{otherwise}}, \\ 
     \end{cases} =
     \begin{cases}
       \mbox{True} &\quad\mbox{\textbf{if}}\hspace{0.25cm}(r>r_{\max}),\\
       \mbox{True} &\quad\mbox{\textbf{if}}\hspace{0.25cm}(r_{\min}<r<r_{\max}) \hspace{0.25cm}\mbox{\textbf{and}}\hspace{0.25cm} \phi'\in \underset{r>\mathcal{R}}{\Xi}(r),\\
       \mbox{False} &\quad\mbox{\textbf{otherwise}}, \\ 
     \end{cases}
\]
with
\[
\underset{r>\mathcal{R}}{\Xi}(r) = \bigcup_{m\in 2\mathbb{N}^{0}+1}^M \Xi_m ,\hspace{0.5cm} \mbox{and}\hspace{0.25cm} \underset{r<\mathcal{R}}{\Xi}(r) = [0,2\pi)\setminus\underset{r>\mathcal{R}}{\Xi}(r).
\]

The inverse distance can be calculated from the previous definitions as
\[
\frac{1}{|\boldsymbol{r}-\boldsymbol{r}'|^\alpha} = 
\sum_{n=0}^{\infty} \sum_{k=0}^{\floor*{n/2}} b_k^{\alpha/2} \xi^{q_k(n)} \left[ \frac{\mathcal{R}^n(\phi')}{r^{n+\alpha}}
\hspace{0.25cm}\mbox{\textbf{if}}\hspace{0.25cm}\vartheta(r,\phi'),\hspace{0.25cm}\mbox{\textbf{else}}\hspace{0.25cm} \frac{r^n}{\mathcal{R}^{n+\alpha}(\phi')} \right].
\]
In the case of the periodically deformed circle $\mathcal{R}(\phi) = R + H f(\phi)$, the binomial theorem and the Taylor series $(1+z)^{-n} = \sum_{s=0}^{\infty}\binom{-n}{s}z^s$ for $|z|<1$ can be introduced to write

\[
\mathcal{R}(\phi)^n = R^n\sum_{s=0}^n \binom{n}{s} \left[\frac{H}{R}f(\phi)\right]^s,
\hspace{0.5cm}\mbox{and}\hspace{0.5cm}  
\frac{1}{\mathcal{R}(\phi)^n} = \frac{1}{R^n}\sum_{s=0}^\infty \binom{-n}{s} \left[\frac{H}{R}f(\phi)\right]^s,
\hspace{0.5cm}\mbox{with}\hspace{0.5cm} H < R.
\]

Hence, the inverse distance can be computed from 
\color{black}
\begin{align}
\frac{1}{|\boldsymbol{r}-\boldsymbol{r}'|^\alpha} =
& \sum_{n=0}^{\infty} \sum_{k=0}^{\floor*{n/2}} b_k^{(\alpha/2)} (\sin\theta)^{q_k(n)}  \left\{ \frac{R^n}{r^{n+\alpha}} \sum_{s=1}^n \binom{n}{s}\left[\frac{H}{R}f(\phi')\right]^s \hspace{0.25cm}\mbox{\textbf{if}}\hspace{0.25cm}\vartheta(r,\phi')\hspace{0.25cm}, \right.\\ & \left.\mbox{\textbf{else}}\hspace{0.25cm} \frac{r^n}{R^{n+\alpha}} \sum_{s=1}^\infty \binom{\alpha-n}{s}\left[\frac{H}{R}f(\phi')\right]^s \right\} [\cos(\phi-\phi')]^{q_k(n)} + \left(\frac{1}{|\boldsymbol{r}-\boldsymbol{r}'|^\alpha}\right)_{circle}, \nonumber
\end{align}
\color{black}
where $|\boldsymbol{r}-\boldsymbol{r}'|_{circle} = \sqrt{r^2 + R^2 - 2rR\sin\theta\cos(\phi-\phi')}$ is the distance to the circle (in the $H=0$ case), and it is obtained by setting the $s$ term of the expansion equal to zero. 
Again, it is possible to implement the Gegenbauer polynomials to write the inverse distance of the circle loop as
\[
\left(\frac{1}{|\boldsymbol{r}-\boldsymbol{r}'|^\alpha}\right)_{circle} = \sum_{n=0}^{\infty} \sum_{k=0}^{\floor*{n/2}} g_{nk}^{(\alpha)}(r,\theta)  [\cos(\phi-\phi')]^{q_k(n)},
\]
where the function $g_{nk}^{(\alpha)}(r,\theta)$ is defined as
\begin{equation}
g_{nk}^{(\alpha)}(r,\theta) := b_k^{(\alpha/2)} (\sin\theta)^{q_k(n)} \left(\frac{R^n}{r^{n+\alpha}} \hspace{0.25cm}\mbox{\textbf{if}}\hspace{0.25cm} r > R, \hspace{0.25cm}\mbox{\textbf{else}}\hspace{0.25cm} \frac{r^n}{R^{n+\alpha}}\right).
\label{gFunctionEq}
\end{equation}

Hence, the inverse distance of the circular-deformed wire can be calculated from the following expansion 

\color{black}
\begin{equation}
\boxed{
\frac{1}{|\boldsymbol{r}-\boldsymbol{r}'|^\alpha}= \sum_{n=0}^{\infty} \sum_{k=0}^{\floor*{n/2}} g_{nk}^{(\alpha)}(r,\theta) [\cos(\phi-\phi')]^{q_k(n)}
    \begin{cases}
       \sum\limits_{s=0}^n \binom{n}{s} \left[\frac{H}{R}f(\phi')\right]^s &\quad\mbox{\textbf{if}}\hspace{0.25cm}\vartheta(r,\phi')\hspace{0.25cm}\mbox{\textbf{is true}},\vspace{0.25cm}\\
        \sum\limits_{s=0}^\infty \binom{-\alpha-n}{s} \left[\frac{H}{R}f(\phi')\right]^s &\quad\mbox{\textbf{otherwise}}, \\ 
     \end{cases}
     }
     \label{inverserExpansionEq}
\end{equation}   
\color{black}
where the expression for the circular case is in the term $s=0$. 
If $r \in \mathbb{R}^{+} \setminus [r_{\min},r_{\max}]$, then there are no roots $\mbox{dim}(\Omega(r))=0$ and the condition $\vartheta(r,\phi')$ is replaced by the simpler $r>r_{\max}$ conditional.

\subsection{Magnetic field in the $r \in \mathbb{R}^{+} \setminus [r_{\min},r_{\max}]$ region}
The radial component of the magnetic field is given from
\[
B_r(\boldsymbol{r}) = \frac{\mu_o i}{4\pi} \cos\theta  \int_0^{2\pi} \frac{ [R^2 + 2RH f(\phi') + H^2 f(\phi')^2]}{|\boldsymbol{r}-\boldsymbol{r}'|^3} d\phi' .
\]

Using the expansion definition for the inverse distance in Eq.~(\ref{inverserExpansionEq}), the $\phi$-component of the magnetic field can be calculated for $r>r_{\max}$ from

\begin{align}
B_r(\boldsymbol{r}) = \frac{\mu_o i}{4\pi} \cos\theta & \sum_{n=0}^{N\rightarrow\infty} \sum_{k=0}^{\floor*{n/2}} g_{nk}^{(\alpha)}(r,\theta) \sum_{s=0}^n \binom{n}{s} \left(\frac{H}{R}\right)^s \nonumber \\ & \int_0^{2\pi} \left\{ [R^2 f(\phi')^s + 2HR f(\phi')^{s+1}+H^2f(\phi')^{s+2}] \right\}  [\cos(\phi-\phi')]^{q_k(n)} d\phi' , \nonumber
\end{align}
Eventually, the infinite sum in the previous formula can be truncated until a finite $N$ to evaluate the magnetic field. 
A similar expression can be obtained for $r<r_{\max}$, and both cases can be written as follows
\begin{equation}
    B_r(\boldsymbol{r}) = \frac{\mu_o i}{4\pi} \cos\theta \sum_{n=0}^{N\rightarrow\infty} \sum_{k=0}^{\floor*{n/2}} g_{nk}^{(3)}(r,\theta) \sum_{s=0}^{\mathcal{M}} \binom{\eta(r,n)}{s} \left(\frac{H}{R}\right)^s \tilde{C}_{nks}(\phi)
\label{BrExpansion1Eq}    
\end{equation}
where $\eta(r,n) = -\alpha-n \hspace{0.1cm}\mbox{\textbf{if}}\hspace{0.1cm}r<r_{\min},\hspace{0.1cm}\mbox{\textbf{else}}\hspace{0.1cm}n$, and $\mathcal{M} = N_2\rightarrow\infty \hspace{0.1cm}\mbox{\textbf{if}}\hspace{0.1cm}r<r_{\min},\hspace{0.1cm}\mbox{\textbf{else}}\hspace{0.1cm}n$. 
The function $\tilde{C}_{nks}(\phi)$ is defined as follows
\begin{equation}
\tilde{C}_{nks}(\phi) := R^2 J_{q_k(n),0}[f^s] + 2 R H J_{q_k(n),0}[f^{s+1}] + H^2 J_{q_k(n),0}[f^{s+2}], 
\label{CTildeCoeffEq}
\end{equation}
where $J_{m,\lambda}[F](\phi)$ are integrals depending on $\phi$ in the sense of
\begin{equation}
J_{m,\lambda}[F](\phi) := \int_{0}^{2\pi} F(\phi')\cos^m(\phi-\phi')\sin^{\lambda}(\phi-\phi')d\phi',
\label{JIntegralDefinitionEq}
\end{equation}
where $m \in \mathbb{N}^0$, and $\lambda$ can be 0 or 1. 
Note that Eq.~(\ref{BrExpansion1Eq}) implicitly contains the solution of the circular loop.
This can be observed by considering the expansion terms that vanish when $H$ is zero. There exists some exceptions (terms with $s=0$), for which this condition implies that
\[
    \lim_{H \to 0} B_r(\boldsymbol{r}) = \frac{\mu_o i}{4\pi} \cos\theta \sum_{n=0}^{N\rightarrow\infty} \sum_{k=0}^{\floor*{n/2}} g_{nk}^{(3)}(r,\theta)  \binom{\eta(r,n)}{0}  (R^2 J_{q_k(n),0}[1]) = \left(B_r(\boldsymbol{r})\right)_{circle}.
\]
Thus, the general solution for the deformed loop can be written more appropriately as follows

\begin{equation}
    B_r(\boldsymbol{r}) = \left(B_r(\boldsymbol{r})\right)_{circle} + \frac{\mu_o i}{4\pi} \cos\theta \sum_{n=0}^{N\rightarrow\infty} \sum_{k=0}^{\floor*{n/2}} g_{nk}^{(3)}(r,\theta) \sum_{s=0}^{\mathcal{M}} \binom{\eta(r,n)}{s} \left(\frac{H}{R}\right)^s C_{nks}(\phi),
\label{BrExpansion2Eq}    
\end{equation}

with 
\[
C_{nks}(\phi) := \mathcal{U}_s R^2 J_{q_k(n),0}[f^s] + 2 R H J_{q_k(n),0}[f^{s+1}] + H^2 J_{q_k(n),0}[f^{s+2}],
\]
and $\mathcal{U}_s = 1-\delta_{s,0}$. The function $\mathcal{U}_s$ was introduced in order to extract the solution of the circular case from the expansion formula\footnote{The function $\mathcal{U}_s$ is necessary because not all the \textit{$s=0$ terms coming from Eq.~\ref{BrExpansion1Eq} provide the magnetic field of the circular current loop} since there are $s=0$ terms in this equation that contribute to the deformed wire solution.}. Such procedure can be used also for the $\theta$-component but it is not necessary for $B_{\phi}$ since $(B_{\phi})_{circle} = 0$. An analogous procedure is used for the $\theta$ and $\phi$ components of the magnetic field, whose results leads to the following vector-form expression
\begin{equation}
\boxed{
\boldsymbol{B}(\boldsymbol{r}) = \boldsymbol{B}_{circle}(\boldsymbol{r}) + \frac{\mu_o i}{4\pi} \sum_{n=0}^{N\rightarrow\infty} \sum_{k=0}^{\floor*{n/2}} g_{nk}^{(3)}(r,\theta) \sum_{s=0}^{\mathcal{M}} \binom{\eta(r,n)}{s} \nu^s \boldsymbol{S}_{nks},
}
\label{expansionFormulaFinalEq}
\end{equation}
where the magnetic field of the circle $\boldsymbol{B}_{circle}(\boldsymbol{r})$ is given by equations (\ref{magneticcircle1})-(\ref{magneticcircle3}), the deformation amplitude by $\nu=\frac{H}{R}$, and the spherical-coordinates components of $\boldsymbol{S}_{nks}$ by
\[
(\boldsymbol{S}_{nks})_r = \cos\theta C_{nks}(\phi) \hspace{0.2cm},\hspace{0.2cm} (\boldsymbol{S}_{nks})_\theta = r P_{nks}(\phi) - \sin\theta C_{nks}(\phi), \hspace{0.2cm}\mbox{and}\hspace{0.2cm}(\boldsymbol{S}_{nks})_\phi = -r\cos\theta Q_{nks}(\phi),
\]
respectively, with 
\[
P_{nks}(\phi) := \mathcal{U}_s R J_{q_k(n)+1,0}[f^s] + H J_{q_k(n)+1,0}[f^{s+1}] - H J_{q_k(n),1}[f^s\dot{f}], \hspace{0.5cm}\mbox{and}\hspace{0.5cm}
\]
\begin{equation}
Q_{nks}(\phi) := R J_{q_k(n),1}[f^s](\phi) + H J_{q_k(n),1}[f^{s+1}](\phi) + H J_{q_k(n)+1,0}[f^s\dot{f}](\phi). 
\label{QCoeffEq}
\end{equation}
In practice, the usefulness of the Eq.~(\ref{expansionFormulaFinalEq}) lies in the ability to obtain analytically all the terms of the expansion, which implies solving exactly $J_{m,\lambda}$ integrals for positive integers $m$ and $\lambda=$ 0 or 1. 
This can still be somewhat difficult for a general definition of $f(\phi)$, but less challenging when the deformation is a harmonic function or a linear combination of harmonic functions. The simplest case is the one discussed in Section~\ref{sec:examples}.

\section{Illustrative example: the harmonically deformed wire}
\label{sec:examples}

\begin{figure}[h]
\begin{minipage}{1.0\textwidth}
  \centering
  \includegraphics[width=0.95\linewidth]{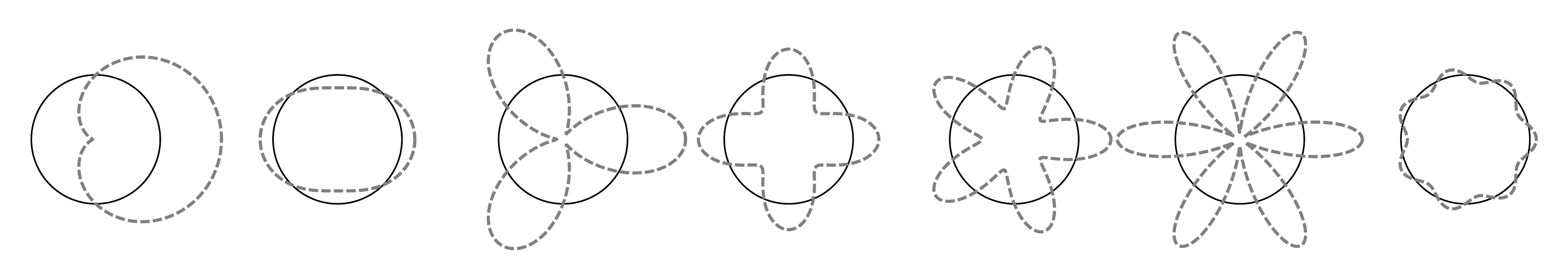}
\end{minipage}
\captionof{figure}{\textcolor{black}{Harmonically deformed curves. Left to right: curves with $p=1,2,\ldots,7$ and $0<H<1$. The unit circle $R=1$ is depicted with a solid line. Note that $p=1$ is the cardioid curve. }}
\label{harmonicallyDeformedWiresFig}
\end{figure}

In the previous sections, the methodology to obtain the magnetic field for a general circular-deformed wire has been explained. As an example case of a particular circular-deformed geometry for the wire, the harmonically deformed curve is studied in the present section. The harmonically deformed curve is defined as $f(\phi) = \cos(p\phi)$, with $p\in\mathbb{N}$. Plots of harmonically deformed curves with several values of $p$ are shown in Fig.~\ref{harmonicallyDeformedWiresFig}. 

Note that if the deformation is harmonic, then
\[
r_{\min} = R (1-\nu) \hspace{0.5cm}\mbox{and}\hspace{0.5cm}   r_{\max} = R (1+\nu), 
\]
and the regions $D_I, D_{II}$, and $D_{III}$ in the $\mathbb{R}^3$ space are defined by $0 \leq r/R < 1-\nu$ , $1-\nu \leq r/R \leq 1+\nu$, and $1-\nu < r/R < \infty$, respectively.

\subsection{Evaluation of integrals for harmonic deformation $f=\cos(p\phi)$}
\label{JIntegralsForHarmonicDeformationAppendixSection}
The integral defined in Eq.~(\ref{JIntegralDefinitionEq}) for harmonic deformation $f=cos(p\phi)$ takes the following form
\[
J_{m,\lambda}[\dot{f}f^s](\phi) := -p\int_{0}^{2\pi} \cos^s(p\phi')\sin(p\phi')\cos^m(\phi-\phi')\sin^{\lambda}(\phi-\phi')d\phi'.
\]

By introducing the complex variable $\mathcal{z}=\exp(\mathbf{i}\phi')$, then Cauchy's residue theorem can be applied to give
\[
J_{m,\lambda}[\dot{f}f^s](\phi) := -p \oint_{\mathcal{c}} j_{ms\lambda}(\mathcal{z},\phi) d\mathcal{z} = -p 2\pi \mathbf{i} \text{Res}[j_{ms\lambda}(0,\phi)],
\]
with $\mathcal{c}$ the unit circle in the complex plane, and
\[
j_{ms\lambda}(\mathcal{z},\phi) = \frac{\mathcal{z}^{-1}}{2^{s+m+\lambda+1}\mathbf{i}^{\lambda+2}}(\mathcal{z}^p+\mathcal{z}^{-p})^s(\mathcal{z}^p-\mathcal{z}^{-p})\left(e^{\mathbf{i}\phi}\mathcal{z}^{-1}+e^{-\mathbf{i}\phi}\mathcal{z}\right)^m\left(e^{\mathbf{i}\phi}\mathcal{z}^{-1}-e^{-\mathbf{i}\phi}\mathcal{z}\right)^\lambda = \sum_{n=-\infty}^\infty a_n \mathcal{z}^n.
\]
In the last expression, $a_n$ are the coefficients of the Laurent series given by the Cauchy's integral formula
\[
a_n(\phi) = \frac{1}{2\pi \mathbf{i}}\oint_c \frac{j_{ms\lambda}(\mathcal{z},\phi)}{\mathcal{z}^{n+1}} d\mathcal{z},
\]
where $j_{ms\lambda}(\mathcal{z},\phi)$ has a non-simple pole at the origin. 
The $a_{-1}$ coefficient is of special interest since it becomes the residue of $j_{ms\lambda}$, being $\text{Res}[j_{ms\lambda}(0,\phi)] = a_{-1}$. 
This coefficient can be obtained by using the binomial theorem, as follows:
\[
j_{ms\lambda}(\mathcal{z},\phi) = \frac{\mathcal{z}^{-1}}{2^{s+m+\lambda+1}\mathbf{i}^{\lambda+2}} \sum_{k=0}^s\binom{s}{k}\sum_{l=0}^m\binom{m}{l}\sum_{t=0}^\lambda\binom{\lambda}{t}\sum_{\sigma=0}^1\binom{1}{\sigma} (-1)^{\lambda-t+1-\sigma}\mathcal{z}^{2(l_{k\sigma s m p \lambda t}-l)}e^{-\mathbf{i}(2l-m+2t-\lambda)\phi},
\]
with $2l_{k\sigma s m p \lambda t}=p(2k+2\sigma-s-1)+m+\lambda-2t$ and $2(l_{k\sigma s m p \lambda t}-l)+1$ the order of the pole. 
Thus, the residue is given by
\[
\text{Res}[j_{ms\lambda}(0,\phi)] = \frac{1}{2^{s+m+\lambda+1}\mathbf{i}^{\lambda+2}} \sum_{k=0}^s\binom{s}{k}\sum_{t=0}^\lambda\binom{\lambda}{t}\sum_{\sigma=0}^1\binom{1}{\sigma} (-1)^{\lambda-t+1-\sigma} e^{-\mathbf{i}(2(k+\sigma)-s-1)p\phi} 
         \begin{bmatrix}
           m \\
           l_{k\sigma s m p \lambda t}
         \end{bmatrix},
\]
with 
\begin{equation}
\begin{bmatrix}
           m \\
           l
\end{bmatrix} := \binom{m}{l} \hspace{0.25cm}\textbf{if}\hspace{0.25cm} l\in[0,1,\ldots,m], \hspace{0.25cm}\textbf{otherwise}\hspace{0.25cm} 0
\label{conditionatedbinomEq}
\end{equation}
a conditioned binomial coefficient. As a result, the first two integrals for $\lambda=0$ and 1 (which are the only that are required to compute in the evaluation of the magnetic field) can be calculated from
\begin{equation}
J_{m,0}[\dot{f}f^s](\phi) = \frac{-2\pi p}{2^{s+m+1}} \sum_{k=0}^s\binom{s}{k}\sum_{\sigma=0}^1\binom{1}{\sigma}  \begin{bmatrix}
           m \\
           l_{k\sigma s m p 0 0}
         \end{bmatrix} (-1)^{1-\sigma} \sin[(2(k+\sigma)-s-1)p\phi], \hspace{0.5cm}\mbox{and}    
\label{JDotLambdaZeroEq}         
\end{equation}
\begin{equation}
J_{m,1}[\dot{f}f^s](\phi) = \frac{2\pi p}{2^{s+m+2}} \sum_{k=0}^s\binom{s}{k}\sum_{t=0}^1\binom{1}{t}\sum_{\sigma=0}^1\binom{1}{\sigma} \begin{bmatrix}
           m \\
           l_{k\sigma s m p 1 t}
         \end{bmatrix} (-1)^{-t-\sigma} \cos[(2(k+\sigma)-s-1)p\phi].    
\label{JDotLambdaOneEq}        
\end{equation}

The residue theorem can also be used to compute $J_{m,\lambda}[f^s](\phi)$, which results for $\lambda=0$ and $\lambda=1$ are, respectively,
\begin{equation}
J_{m,0}[f^s](\phi) = \frac{2\pi}{2^{s+m}} \sum_{k=0}^s\binom{s}{k} \begin{bmatrix}
           m \\
           l'_{k s m p 0 0}
         \end{bmatrix} \cos[(2k-s)p\phi], \hspace{1cm}\mbox{and}     
\label{JLambdaZeroEq}                  
\end{equation}
\begin{equation}
J_{m,1}[f^s](\phi) = \frac{2\pi}{2^{s+m+\lambda}} \sum_{k=0}^s\binom{s}{k}\sum_{t=0}^1\binom{1}{t} (-1)^{1-t} \begin{bmatrix}
           m \\
           l'_{k s m p 1 t}
         \end{bmatrix} \sin[(2k-s)p\phi] ,
\label{JLambdaOneEq}                           
\end{equation}
with $2l'_{k s m p \lambda t} = p(2k-s)+m+\lambda-2t$.

\subsection{Magnetic field calculation}

The fundamental calculation to obtain the magnetic field via expression (\ref{expansionFormulaFinalEq}) lies in the integral result of Eq.~(\ref{JIntegralDefinitionEq}). These multiple integrals can be evaluated straightforwardly by using standard tools of complex analysis, such as the residue theorem (see Section \ref{JIntegralsForHarmonicDeformationAppendixSection}).  

\begin{figure}[h]
\begin{minipage}{.33\textwidth}
  \centering
  \includegraphics[width=1.0\linewidth]{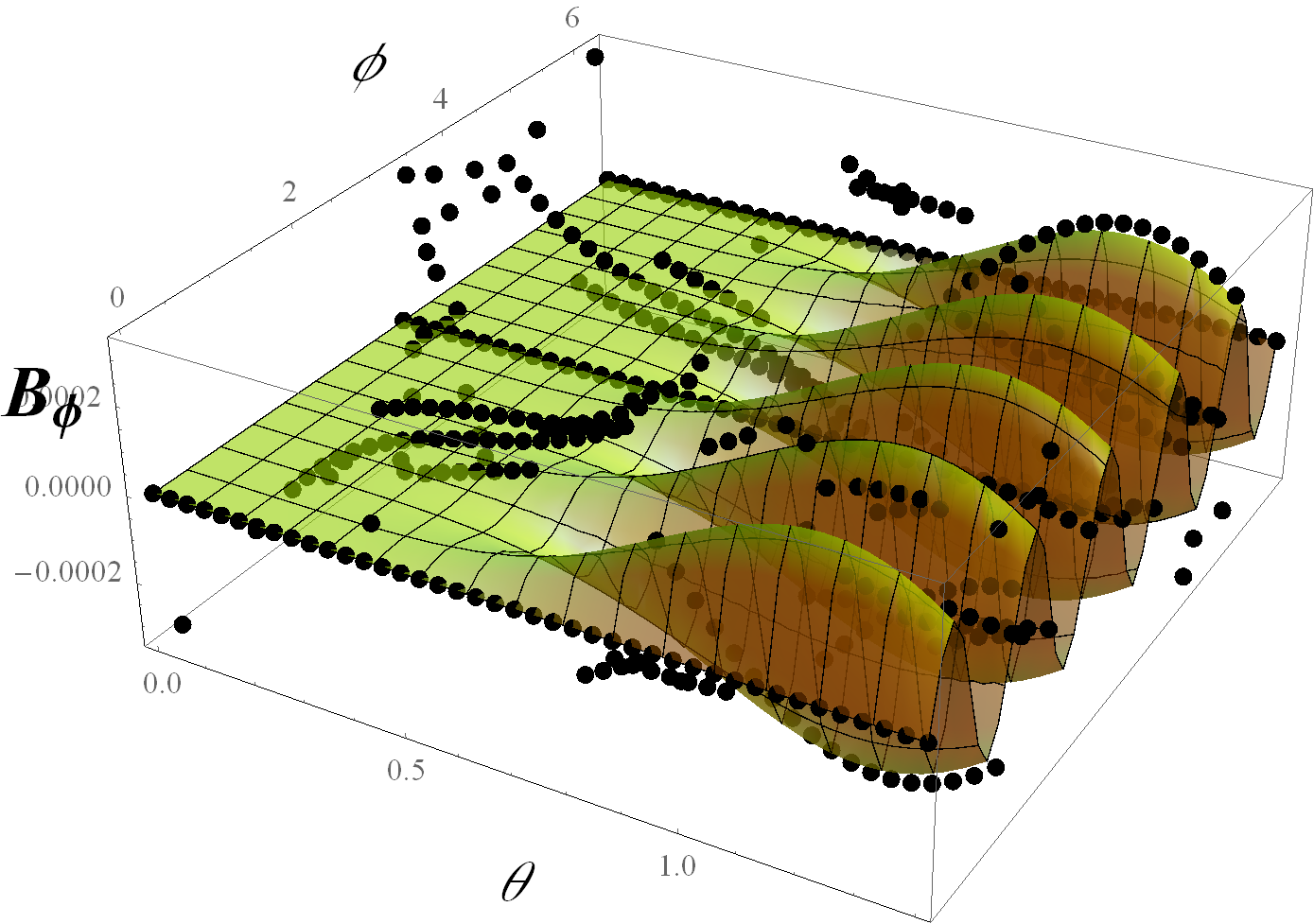}\\
  \caption*{\small(a) 1 digit accuracy, $\tau_{num}=4.98s$  } 
\end{minipage}%
\begin{minipage}{.33\textwidth}
  \centering
  \includegraphics[width=1.0\linewidth]{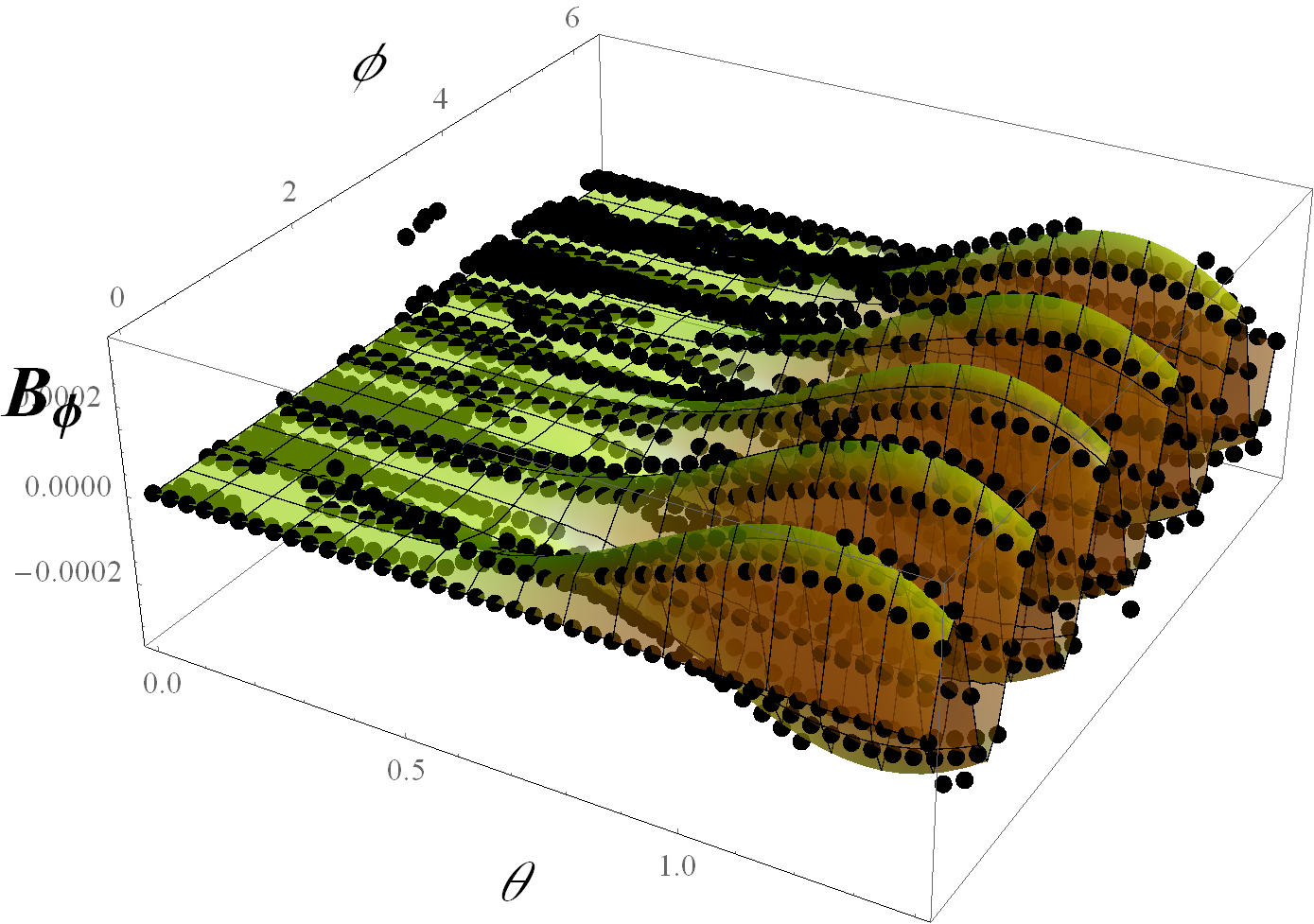}\\
  \caption*{\small(b) 2 digit accuracy, $\tau_{num}=11.15s$ }
\end{minipage}
\begin{minipage}{.33\textwidth}
  \centering
  \includegraphics[width=1.0\linewidth]{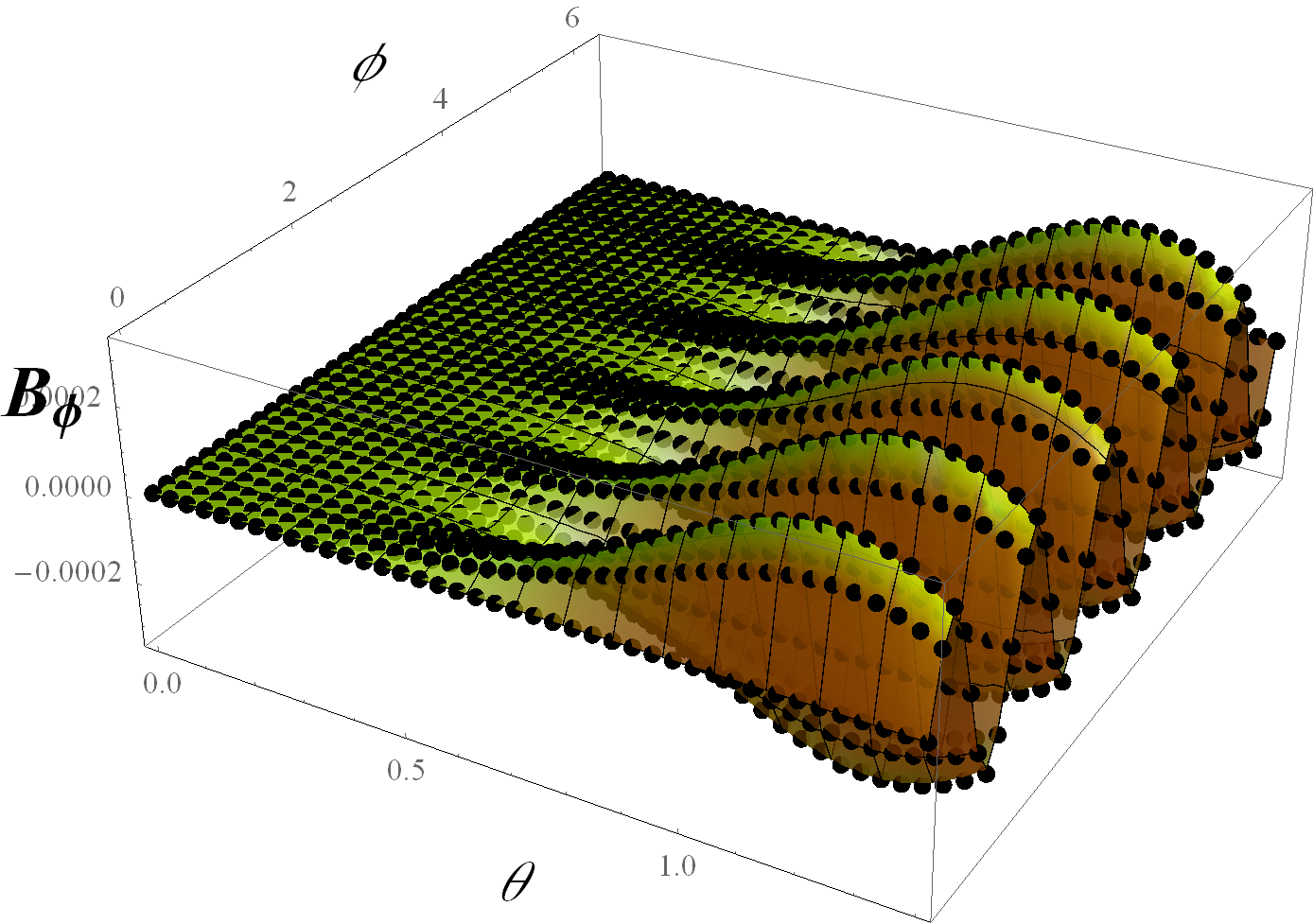}
  \caption*{\small(c) 6 digit accuracy, $\tau_{num}=29.68s$ }
\end{minipage}
\captionof{figure}{Analytical expansion results. The color-plot surface in (a), (b), and (c) corresponds to the solution given by Eq.~(\ref{BphiApproxN9p5nu5Eq}) for $p=5$, $\nu=1/2$, and $R=1$. This surface has been developed over the sphere $S$ of radius $r=1.5r_{\max}$ centered at the origin. The surface is evaluated using 1600 scattered spatial points and it takes $\tau_{expansion} = 0.68s$ to be computed. Results are contrasted with numerical integration (represented by using 1600 black dots): (a) 1 digit accuracy, (b) 2 digit accuracy, and (c) 6 digit accuracy.}
\label{accuracyFig}
\end{figure}

The integrals $J_{m,\lambda}[f^s](\phi)$ and $J_{m,\lambda}[\dot{f}f^s](\phi)$ are linear combinations of harmonic functions.
For example, the $p=5$ curve and $m=31$ problem requires $J_{31,1}[\dot{f}f^{11}](\phi)$, which gives
\begin{align}
J_{31,1}[\dot{f}f^{11}](\phi) & = -p\int_{0}^{2\pi} \cos^{11}(p\phi')\sin(p\phi')\cos^{31}(\phi-\phi')\sin(\phi-\phi')d\phi' \nonumber \\ & =  \frac{4125 \pi  (3225612 \cos (10 \phi )+56637 \cos (20 \phi )+2 \cos (30 \phi ))}{2199023255552}. \nonumber
\end{align}
Once the $J$'s integrals are found analytically, it is possible to build expansion formulas for the magnetic field straightforwardly with (\ref{expansionFormulaFinalEq}). 
For instance, the $\phi$-component given by Eq.~(\ref{expansionFormulaFinalEq}) requires to compute the coefficients defined in  Eq.~(\ref{QCoeffEq}) via the residue theorem (see Appendix Section \ref{JIntegralsForHarmonicDeformationAppendixSection} for a detailed explanation). 

Only for clarifying purposes, the $\phi$-component for the magnetic field given by the expansion approach is demonstrated for the $p=5$ curve and $r>r_{\max}=R+H$, giving 
\begin{align}
B_{\phi}(\boldsymbol{r})  = & - \frac{3465\nu}{262144 r^{10}}\sin ^4(\theta ) \cos (\theta ) \sin (5 \phi ) [1248-512 r^2+8736 \nu ^2-1920 r^2 \nu
   ^2+1625 \nu ^4-520 r^2 \nu ^4 \nonumber \\ & + 260 (8+56 \nu ^2+(15+2 r^2) \nu ^4) \cos (2 \theta )-5525 \nu ^4 \cos (4 \theta )] + O(\nu^6) ,
\label{BphiApproxN9p5nu5Eq}   
\end{align}
where the expansion is truncated up to $N=9$ and fifth-order in $\nu$.

Some other solutions for the harmonically deformed curves with $p=1$ (Cardioid curve) and $p=3$ are presented in the Appendix~\ref{TruncatedAppendixSection}.

\begin{figure}[h]
\begin{minipage}{.33\textwidth}
  \centering
  \includegraphics[width=1.0\linewidth]{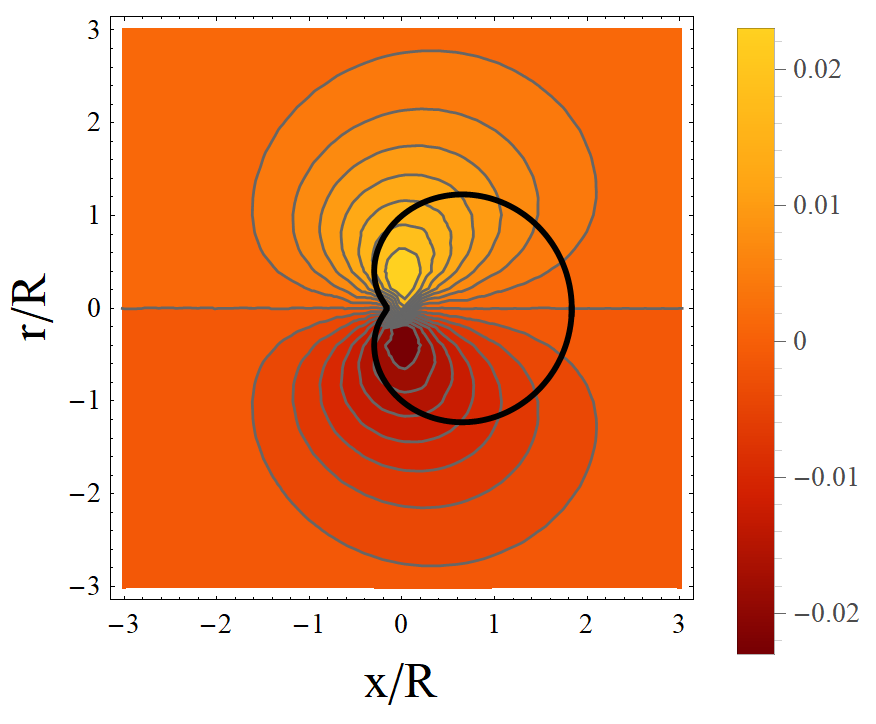}
  \caption*{(a) $p=1,\nu=5/6$ (cardioid) } 
\end{minipage}%
\begin{minipage}{.33\textwidth}
  \centering
  \includegraphics[width=1.0\linewidth]{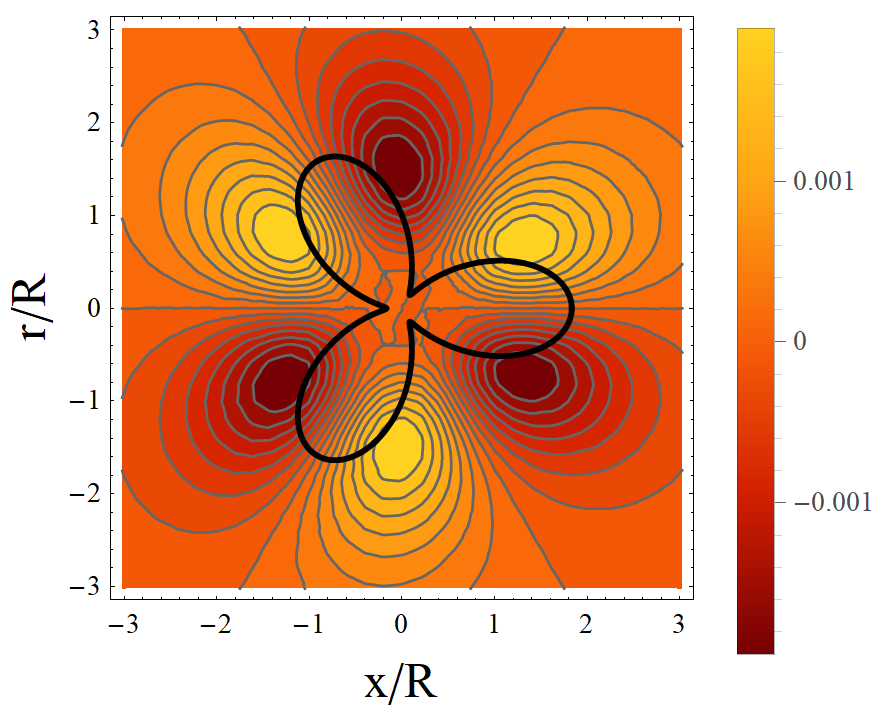}
  \caption*{(b) $p=3,\nu=5/6$}
\end{minipage}
\begin{minipage}{.33\textwidth}
  \centering
  \includegraphics[width=1.0\linewidth]{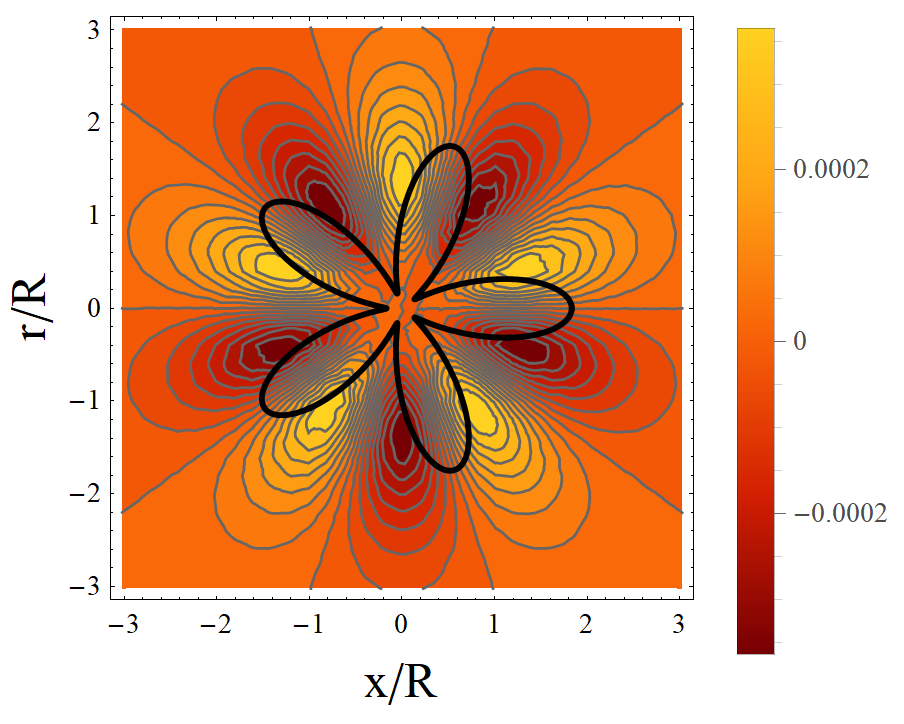}
  \caption*{(c) $p=5,\nu=5/6$}
\end{minipage}

\captionof{figure}{Polar angle component of the magnetic field $B_\phi$ evaluated at the $z=z_o$ plane. All plots correspond to truncated expansions of harmonically deformed curves. }
\label{phiPleuseursFig}
\end{figure}

\begin{figure}[h]
\begin{minipage}{.33\textwidth}
  \centering
  \includegraphics[width=1.0\linewidth]{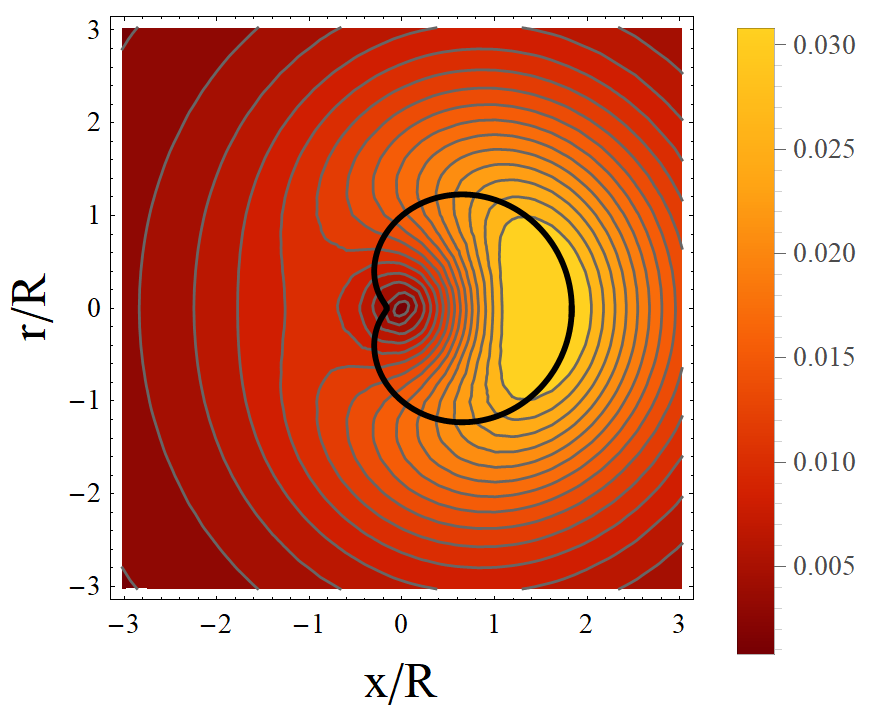}
  \caption*{(a) $\nu=0$ (circle) } 
\end{minipage}%
\begin{minipage}{.33\textwidth}
  \centering
  \includegraphics[width=1.0\linewidth]{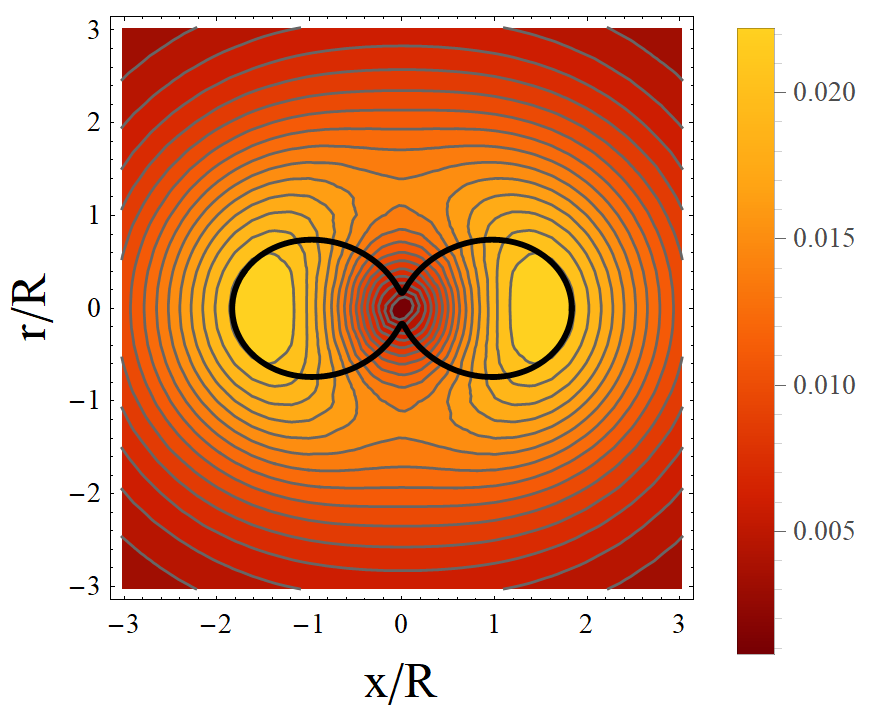}
  \caption*{(b) $\nu=1/5$}
\end{minipage}
\begin{minipage}{.33\textwidth}
  \centering
  \includegraphics[width=1.0\linewidth]{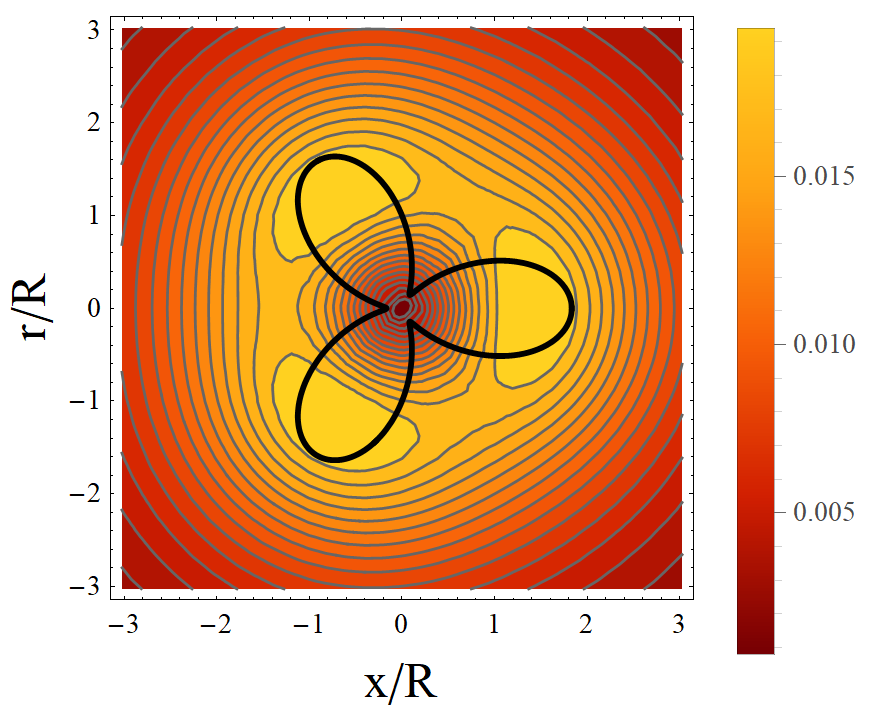}
  \caption*{(c) $\nu=9/10$}
\end{minipage}
\captionof{figure}{Horizontal radial component of the magnetic field $(B_u(\boldsymbol{r}))^{(\nu)} = (B_r(\boldsymbol{r}))^{(\nu)}\sin\theta$ evaluated at the $z=z_o$ plane. Plots correspond to truncated expansions of harmonically deformed curves: (a) $p=1$, (b) $p=2$, and (c) $p=3$.}
\label{uPleuseursFig}
\end{figure}

\begin{figure}[h]
\begin{minipage}{.33\textwidth}
  \centering
  \includegraphics[width=1.0\linewidth]{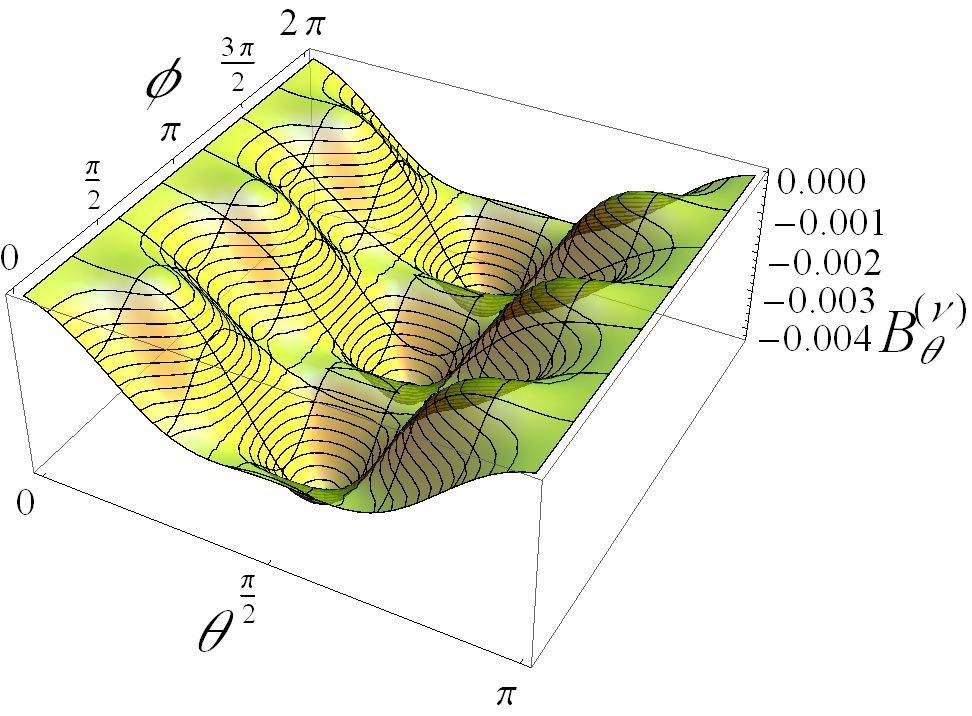}
  \caption*{(a) $p=3$ } 
\end{minipage}%
\begin{minipage}{.33\textwidth}
  \centering
  \includegraphics[width=1.0\linewidth]{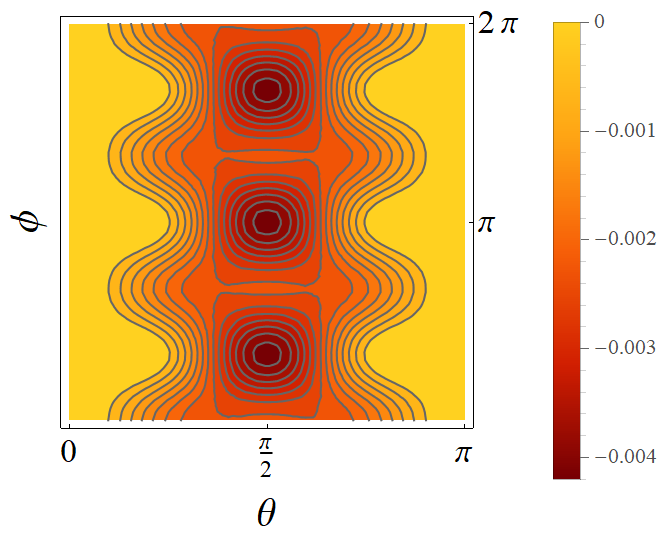}
  \caption*{(b) $\tau_{expansion}=0.78s$}
\end{minipage}
\begin{minipage}{.33\textwidth}
  \centering
  \includegraphics[width=1.0\linewidth]{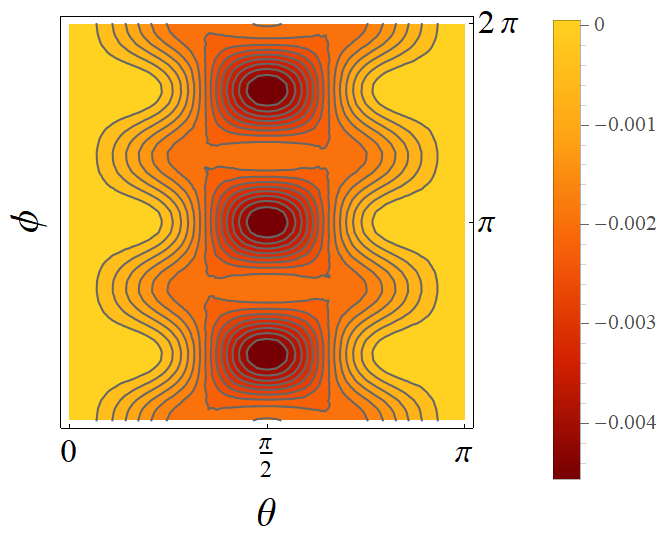}
  \caption*{(c) $\tau_{numeric}=32.21s$}
\end{minipage}
\captionof{figure}{Analytical expansion results of the azimuth component of the magnetic field $(B_{\theta}(\boldsymbol{r}))^{(\nu)}$. The surface in (a) and (b) corresponds to the solution given by the expansion expression with $p=3$ and $\nu=1/10$. This surface is developed over the sphere $S$ of radius $r=r_{\min}/4$ and located in the center of the loop. The evaluation procedure takes $\tau_{expansion} = 0.78s$ to be computed. Analytical results are contrasted with numerical integration in (c), that takes $\tau_{numeric}=32.21s$ to be computed.
}
\label{BThetaNup3Fig}
\end{figure}

A plot of the $B_{\phi}(\boldsymbol{r})$ component of the magnetic field given by Eq.~(\ref{BphiApproxN9p5nu5Eq}) is shown in Fig.~\ref{accuracyFig}.
The plot has been presented for the developed surface over the sphere of radius $r=3r_{\max}/2$  centered at the origin. Expansion formulas of $B_\phi$ for other $p$ curves are not presented here, but those can be obtained with the same procedure. Plots of analytical results for the polar angle component on a $z=1.25 r_{\max}$ plane are shown in Fig.~\ref{phiPleuseursFig}. As expected, the $p$ oscillations of the loop are well represented by $B_{\phi}(\boldsymbol{r})$. Other magnetic field component plots for the harmonically deformed curves are presented in Fig.~\ref{uPleuseursFig} and \ref{BThetaNup3Fig}. In the case of the horizontal radial component of the magnetic field for $p=1,2$ and $3$, these results are shown in Fig.~\ref{uPleuseursFig}.

For the sake of comparison, the magnetic field can be also obtained from the numerical integration of the Biot-Savart formula. 
Discrete results in Fig.~\ref{accuracyFig} are represented by using black dots and correspond to the application of the numerical integration using a closed Newton-Cotes rule of third-order (or Simpson's 3/8 rule) in the same problem set.
The NIntegrate routine in Mathematica \cite{wolfram2012version} has been applied to perform the numerical integration.
This function has been tested using different values of the AccuracyGoal option: a parameter that defines the accuracy of the numerical integration.
An AMD Ryzen 31200 Quad-Core Procesor 3.10GHz with 8GB of RAM has been used both to evaluate the expansion formula in the spatial domain and to perform the numerical integration of the Biot-Savart problem. A serial computation has been performed in both calculations.
Indeed, expression in Eq.~(\ref{BphiApproxN9p5nu5Eq}) has been also written in Mathematica to draw the surface in Fig.~\ref{accuracyFig}.
This surface has been constructed by evaluating 1600 scattered spatial points in Eq.~(\ref{BphiApproxN9p5nu5Eq}), which has taken about $\tau_{expansion} = 0.68s$ to complete the serial computation.

\begin{figure}[h]
\begin{minipage}{.33\textwidth}
  \centering
  \includegraphics[width=0.8\linewidth]{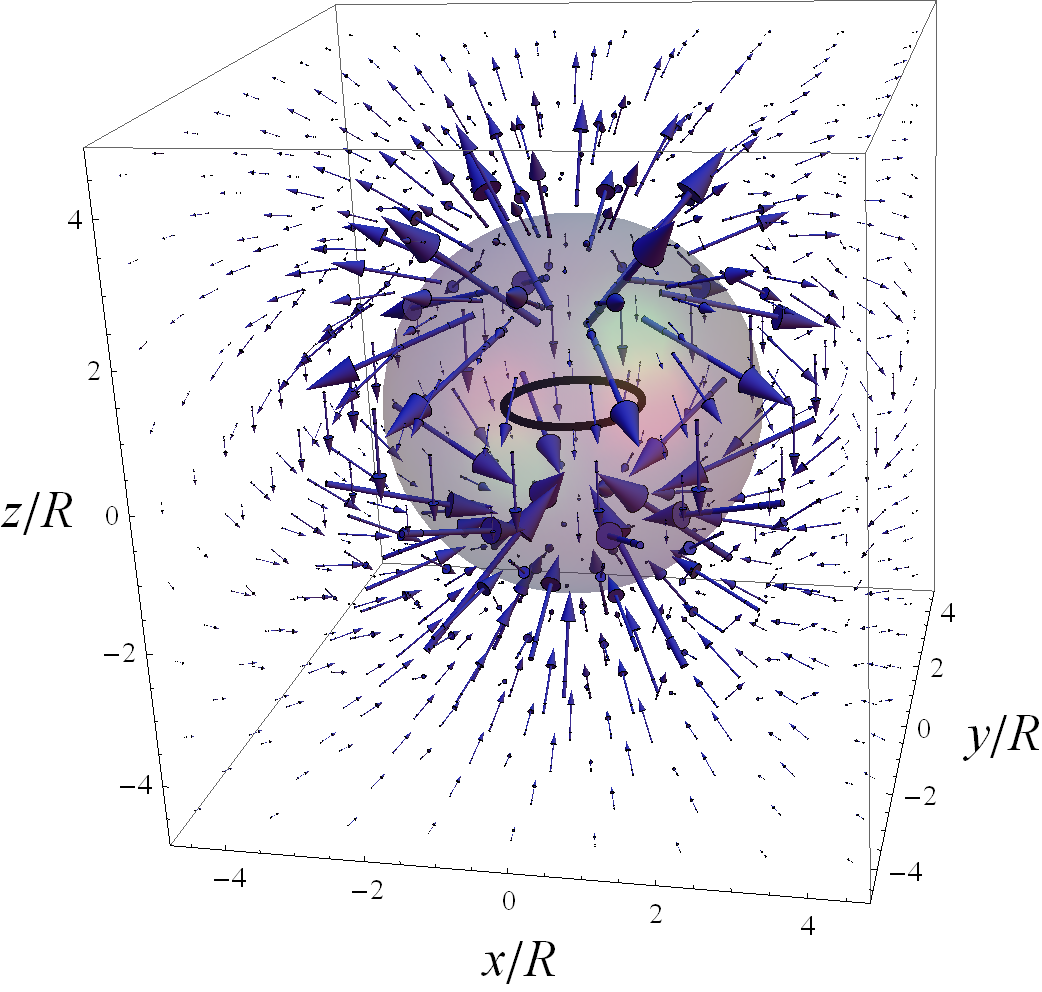}\\
  \includegraphics[width=0.9\linewidth]{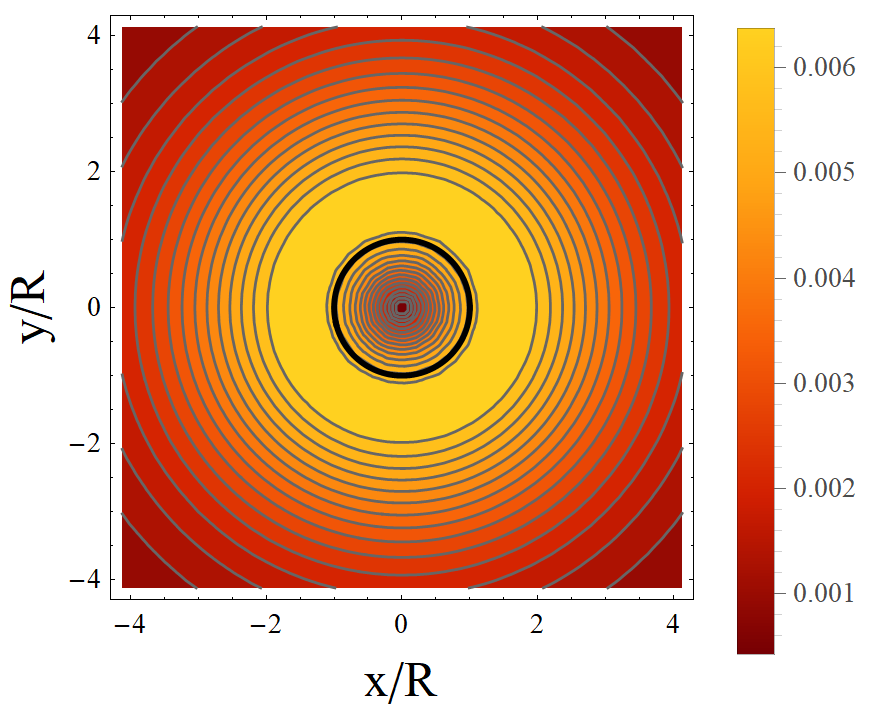}
  \caption*{(a) $\nu=0$ (circle) } 
\end{minipage}%
\begin{minipage}{.33\textwidth}
  \centering
  \includegraphics[width=0.8\linewidth]{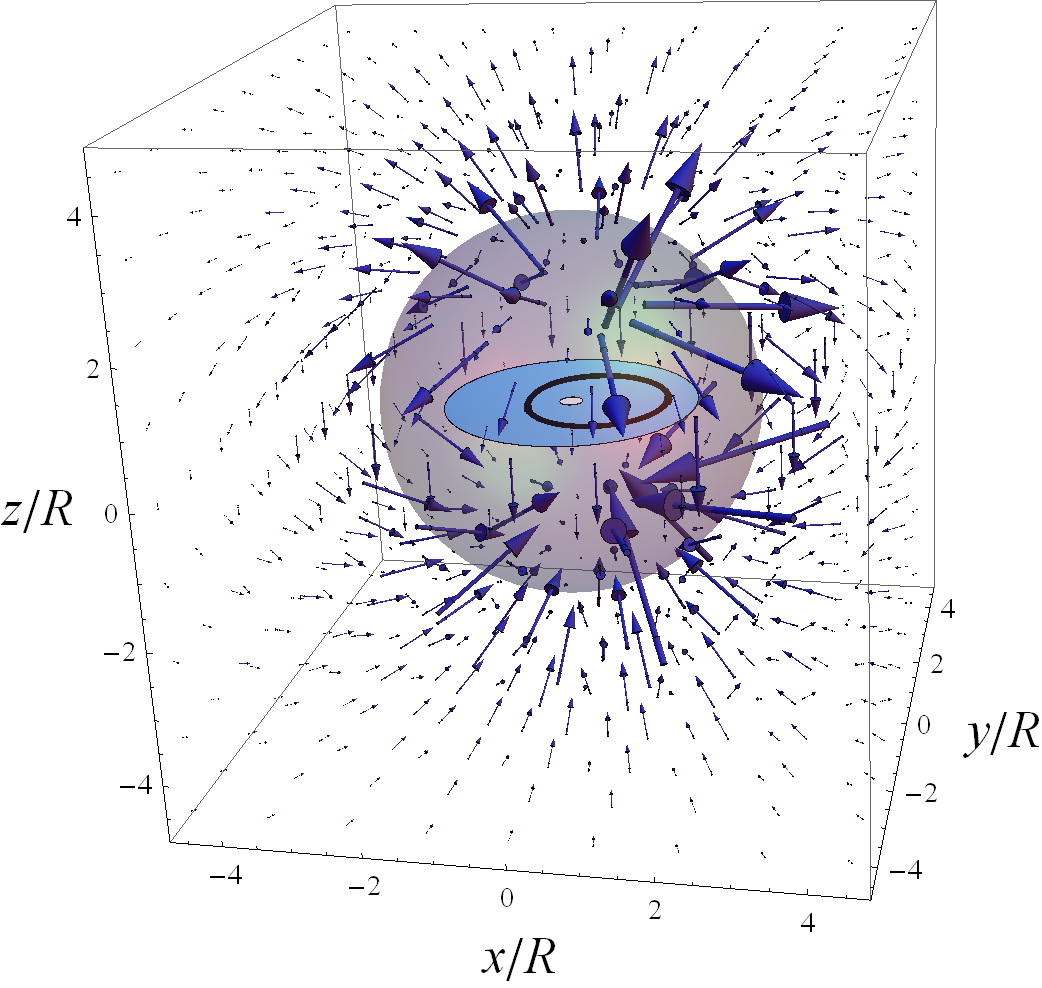}\\
  \includegraphics[width=0.9\linewidth]{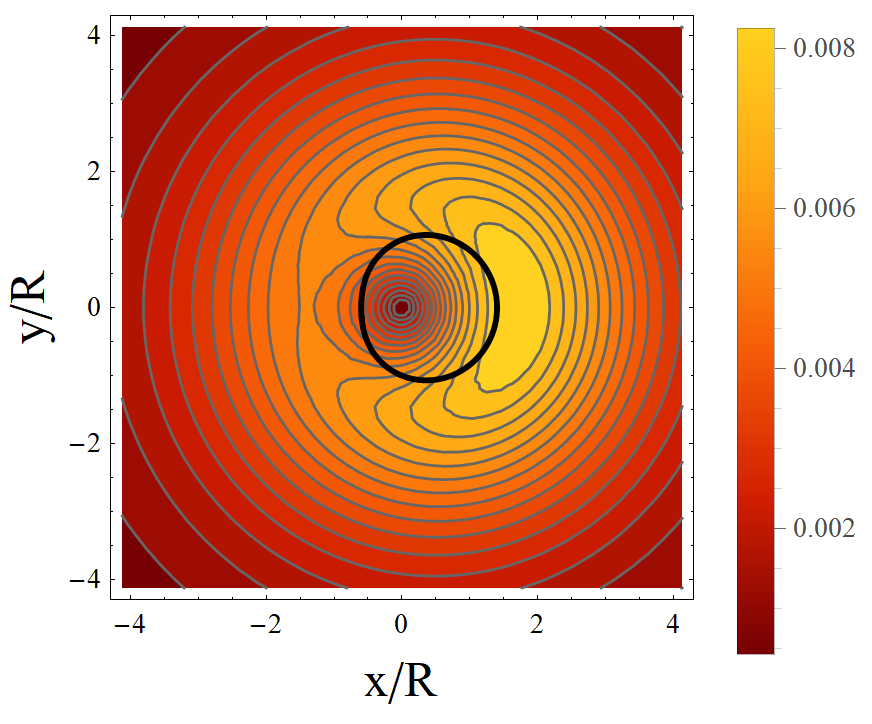}
  \caption*{(b) $\nu=1/5$}
\end{minipage}
\begin{minipage}{.33\textwidth}
  \centering
  \includegraphics[width=0.8\linewidth]{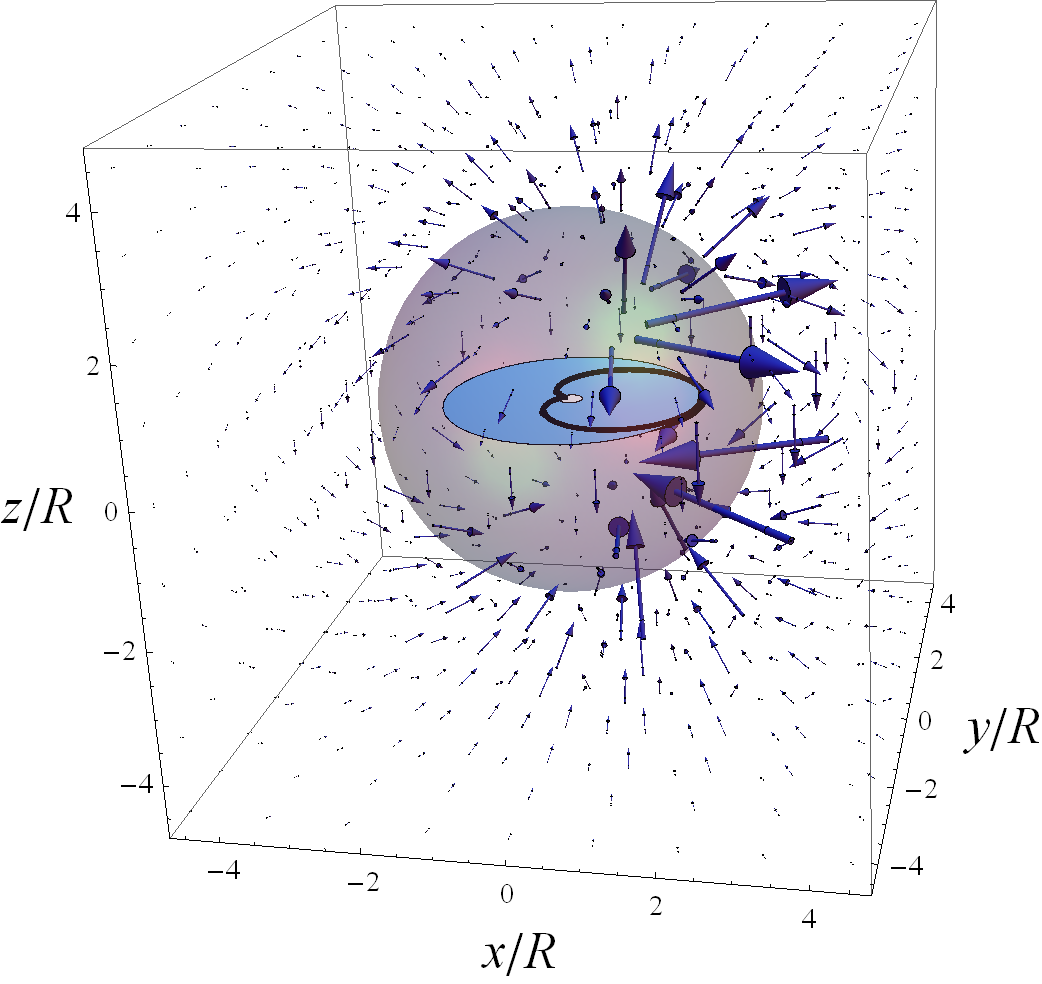}
  \includegraphics[width=0.9\linewidth]{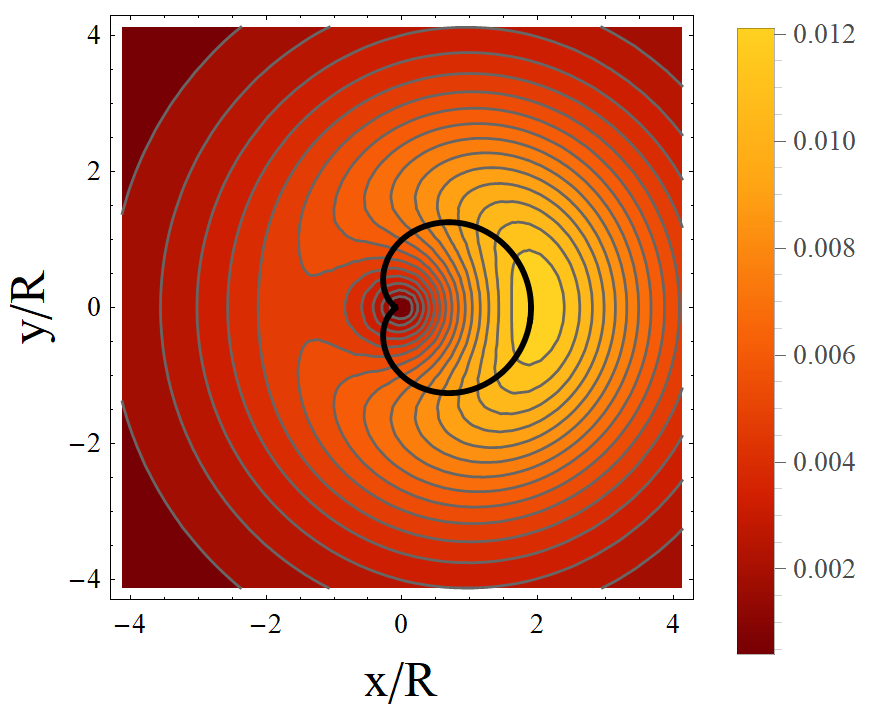}
  \caption*{(c) $\nu=9/10$}
\end{minipage}
\captionof{figure}{Analytical results of the cardioid curve. (top) Magnetic vector field depicted outside of the sphere S of radius $3r_{\max}/2$ and (bottom) horizontal radial component of the magnetic field $B_u(\boldsymbol{r}) = B_r(\boldsymbol{r})\sin\theta$ evaluated on the $z=3r_{\max}/2$ plane that cuts the north pole of S. Cardioid loop with (a) $\nu=H/R=0$, (b) $\nu=H/R=1/5$, and (c) $\nu=H/R=9/10$.}
\label{cardioidFig}
\end{figure}

Finally, Fig.~\ref{cardioidFig} shows the magnetic field for a circular-deformed wire in the form of the cardioid loop $p=1$. 
Analytical results of the magnetic vector field are depicted over the sphere S of radius $3r_{\max}/2$ at the top of that figure.
The vector fields in Fig.~\ref{cardioidFig} correspond to truncated expansions of Eq.~(\ref{expansionFormulaFinalEq}) presented in Appendix \ref{TruncatedAppendixSection}.
Physical behavior can be identified in all deformation wire cases.
Also, the horizontal radial component of the magnetic field $B_u(\boldsymbol{r}) = B_r(\boldsymbol{r})\sin\theta$ evaluated on the $z=3r_{\max}/2$ plane that cuts the north pole of S is presented at the bottom of the same figure.
Deviant behavior can be identified according to the degree of deformation of the circular wire.

\color{black}
In general, the magnetic field given by Eq.~(\ref{expansionFormulaFinalEq}) and the standard dipolar approximation should coincide if $\boldsymbol{B}$ is evaluated far from the wire $r/R\gg1$. The magnetic field according to the dipolar approximation is
\begin{equation}
    B_r^{dip}(r,\theta) = \frac{\mu_o m}{2\pi r^3}\cos\theta,  \hspace{0.5cm} B_\theta^{dip}(r,\theta) = \frac{\mu_o m}{4\pi r^3}\sin\theta,  \hspace{0.5cm} \mbox{and} \hspace{0.5cm} B_\phi^{dip}(r,\theta) = 0    
\end{equation}

where $\mathcal{m}$ is the magnetic moment. In the current problem, this parameter can be written as follows
\[
\boldsymbol{\mathcal{m}} = \pi i R^2 ( 1 + 2\nu \langle f \rangle + \nu^2 \langle f^2 \rangle) \hat{z},  \hspace{0.5cm}\mbox{with}\hspace{0.5cm} \langle f \rangle := \frac{1}{2\pi}\int_0^{2\pi} f(\phi) d\phi 
\]
the average of the deformation function $f(\phi)$ in $[0,2\pi)$. Again, if we consider the harmonic deformation function $f(\phi)=\cos(p\phi)$, then the magnetic moment gives $\mathcal{m}=\pi i R^2 (1+\nu^2/2)$ since $\langle f \rangle=0$ and $\langle f^2 \rangle =1/2$. 

\begin{figure}[h]
\begin{minipage}{.33\textwidth}
  \centering
  \includegraphics[width=1.0\linewidth]{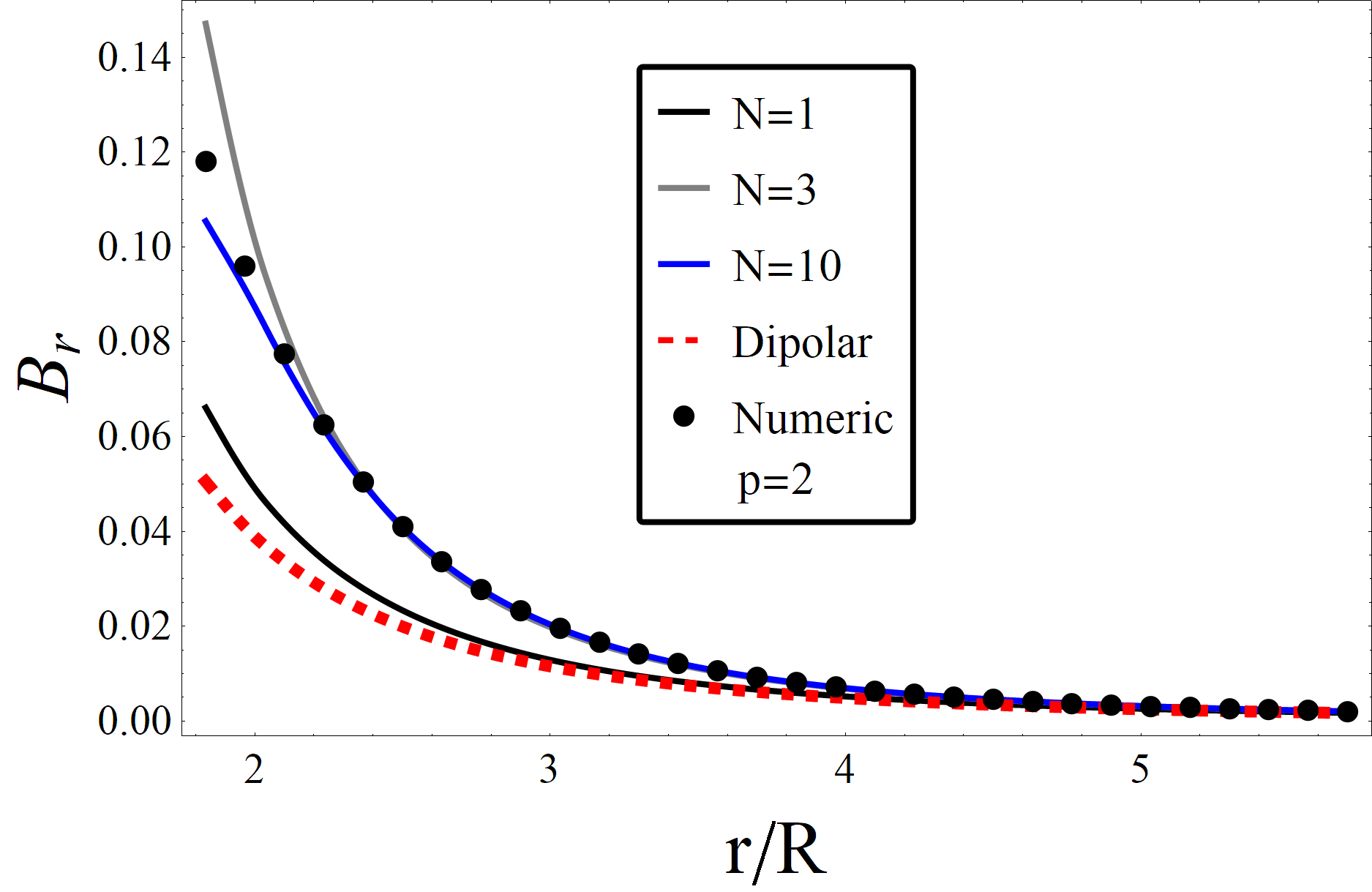}\\
  \caption*{\small(a)  $B_r$} 
\end{minipage}%
\begin{minipage}{.33\textwidth}
  \centering
  \includegraphics[width=1.0\linewidth]{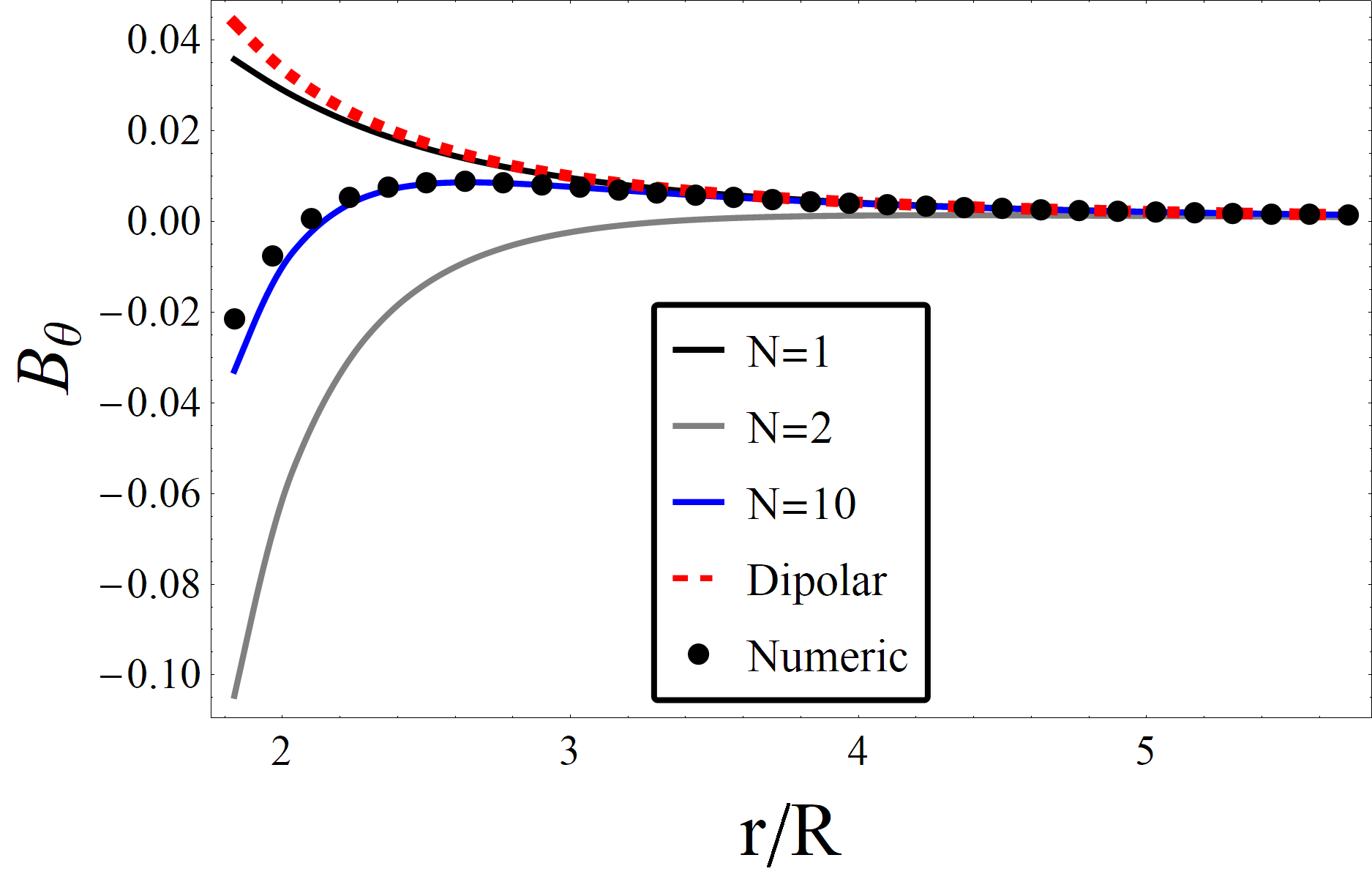}\\
  \caption*{\small(b) $B_\theta$ }
\end{minipage}
\begin{minipage}{.33\textwidth}
  \centering
  \includegraphics[width=1.0\linewidth]{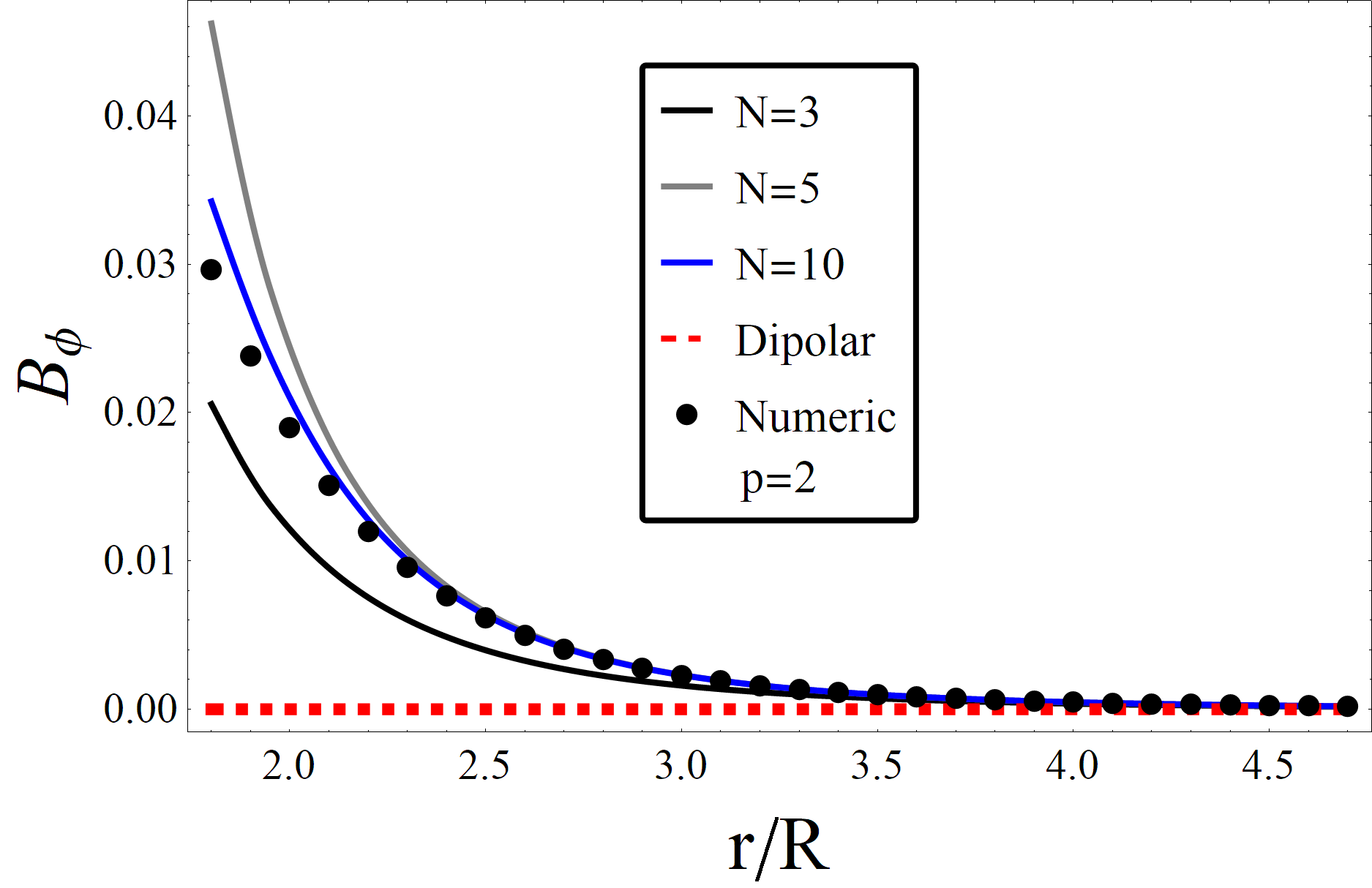}
  \caption*{\small(c) $B_\phi$ }
\end{minipage}
\captionof{figure}{Analytical expansion solution and dipolar approximation for $p=2$ and $\nu=0.7$. The plots in the region $\mathcal{D}_{\text{III}}\subset\mathbb{R}^3$ are defined within $1.7<r/R<\infty$. The angular coordinates are set as $\theta=\pi/4$ and $\phi=\pi/18$ in all components of the magnetic field. Black points correspond to numerical integration, the red dashed line to the dipolar approximation, and solid lines to the $N$-finite expansion in Eq.~(\ref{expansionFormulaFinalEq}). }
\label{dipolarFig}
\end{figure}

Therefore, the dipolar magnetic field does not have a dependence on $p$ and grows quadratically with the deformation parameter $\nu$. A comparison among the dipolar approximation, truncated expansions of Eq.~(\ref{expansionFormulaFinalEq}) for several values of $N$, and the numerical results of the magnetic field are shown in Fig.~\ref{dipolarFig}. In that figure the magnetic field due to a deformation function $f(\phi)=\cos(2\phi)$ with $\nu=0.7$ is evaluated by setting $\theta=\pi/2$ and $\phi=\pi/18$. It can be observed in Fig.~\ref{dipolarFig} that the magnetic fields from different truncations converge to the numerical integration as $r/R$ grows and the convergence is better if the expansion include more terms. On the other hand, the dipolar approximation converge to the expansion and numerical results for large values of $r/R$. The dipolar approximation and the Eq.~(\ref{expansionFormulaFinalEq}) including few terms (this $N=1,2$ or 3) tend to fail (in general) if $r/R$ is not sufficiently large. The result from Eq.~(\ref{expansionFormulaFinalEq}) can be refined by increasing $N$, for instance until $N=10$. Naturally, there persists deviations between Eq.~(\ref{expansionFormulaFinalEq}) and the numerical integration as $r \rightarrow r_{\max}^{+}$ (exactly, as $r/R \rightarrow 1.7^{+}$), since the point of evaluation approaches to the $D_{II}$ region where the wire lies.

\subsubsection{Error estimates}
To deepen in the analysis, the following $L^2-$relative error norm is calculated over the magnetic field results,
\[
\Xi = \sqrt{\frac{1}{\sum_{(r_i,\theta_j,\phi_k)\in\mathcal{P}}  |\boldsymbol{B}( \boldsymbol{r}_{ijk} )|^2}\sum_{(r_i,\theta_j,\phi_k)\in\mathcal{P}} |\Delta \boldsymbol{B}( \boldsymbol{r}_{ijk} )|^2 }.
\]
Here $\Delta \boldsymbol{B}(\boldsymbol{r}) = \boldsymbol{B}(\boldsymbol{r}) - \boldsymbol{B}^{(\text{num})}(\boldsymbol{r})$, where $\boldsymbol{B}(\boldsymbol{r})$ and $\boldsymbol{B}^{(\text{num})}(\boldsymbol{r})$ are the magnetic field computed with Eq.~(\ref{BrExpansion1Eq}) and via numerical integration, respectively.

\begin{figure}[h]
  \centering
  \includegraphics[width=0.8\linewidth]{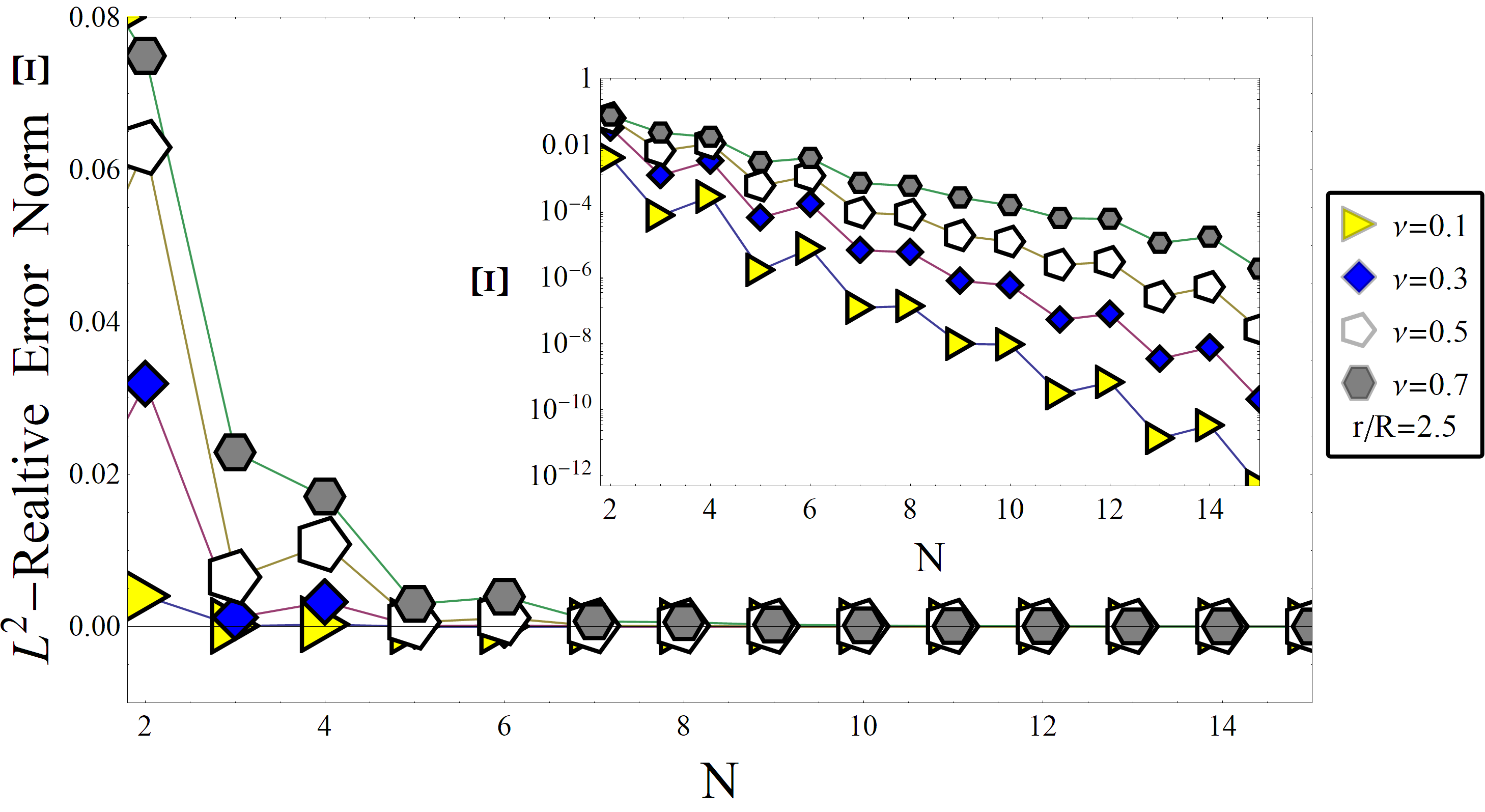}
\captionof{figure}{{\color{black}$L^2-$relative error norm of $\boldsymbol{B}$ as a function of the expansion truncation.}}
\label{relativeErrorNormAFig}
\end{figure}

Fig.~\ref{relativeErrorNormAFig} shows the $L^2-$relative error norm convergence against the $N$-th integer of the expansion truncation. In this case, the three components of the magnetic field have been evaluated at $r=2.5R \in \mathcal{D}_{\text{III}}$ for a fixed harmonic deformation $f(\phi)=\cos(2\phi)$ and several different deformation amplitudes $\nu$. Analytic results are computed using Eq.~(\ref{expansionFormulaFinalEq}) by truncating the first sum of the expansion up to a $N$ integer. The reference set $\mathcal{P}$ (used to compute the $L^2$-relative errors) is constructed by numerically evaluating the solution over a thousand nodes laying on the $\left\{ (r,\theta,\phi) | \theta\in[0,\pi) \wedge \phi\in[0,2\pi) \right\}$ hemisphere of radius $r$. It can be observed that the error is large for small values of $N$ since the expansion includes few terms. As the value of N increases, the analytical expansion solution gets closer to the numerical one and the error approaches to zero. The amplitude of the deformation $\nu\in[0,1)$ generally affects the error of the analytical solution: as $\nu$ grows, more terms are required in the expansion to reproduce the solution.



On one hand, if $r/R \rightarrow (1+\nu)^{+}$, then there are points $\boldsymbol{r} \in D_{III}$ located on the hemisphere $\mathcal{P}$ which are in the neighborhood of the wire.
On the other hand, for the inner spherical region $D_I$ of radius $r_{\min}=R(1-\nu)$, the error increases as $r$ approaches to $r_{\min}$ and decreases as it tends to the origin.

\begin{figure}[H]
\centering
\begin{minipage}{.33\textwidth}
  \centering
  \includegraphics[width=0.6\linewidth]{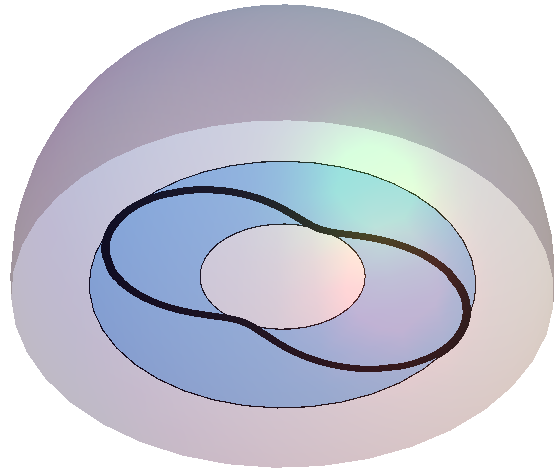}
  \caption*{(a) $r=1.4r_{max}$}
\end{minipage}
\begin{minipage}{.33\textwidth}
  \centering
  \includegraphics[width=0.6\linewidth]{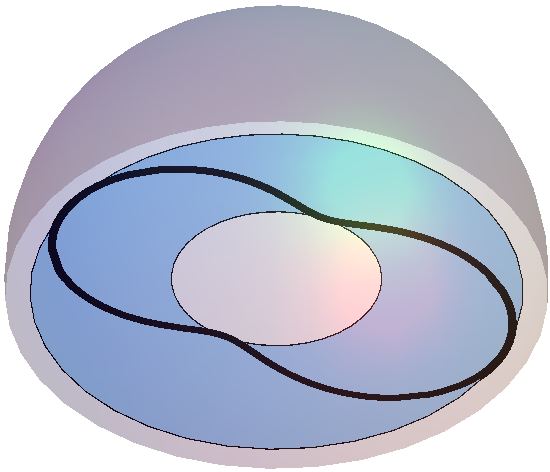}
  \caption*{(b) $r=1.1r_{max}$}
\end{minipage}

\begin{minipage}{.33\textwidth}
  \centering
  \includegraphics[width=1.0\linewidth]{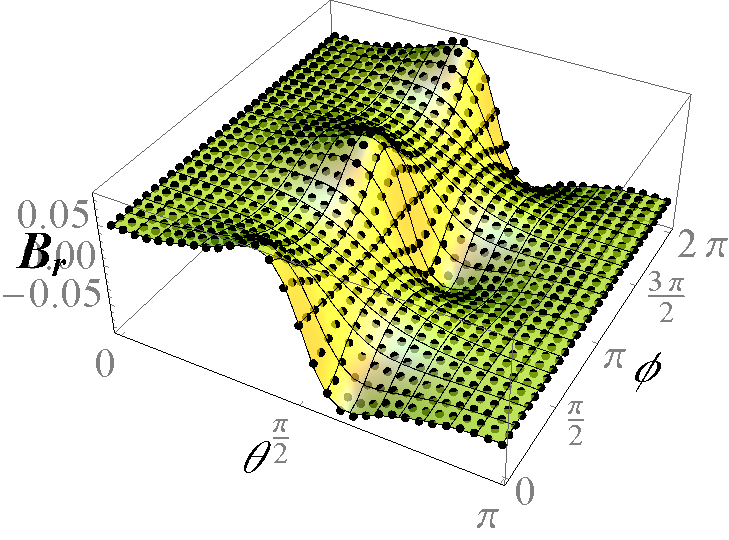}
  \caption*{(c) $r=1.4r_{max}, \frac{\tau_{num}}{\tau_{exp}} = 5.79$ } 
\end{minipage}%
\begin{minipage}{.33\textwidth}
  \centering
  \includegraphics[width=1.0\linewidth]{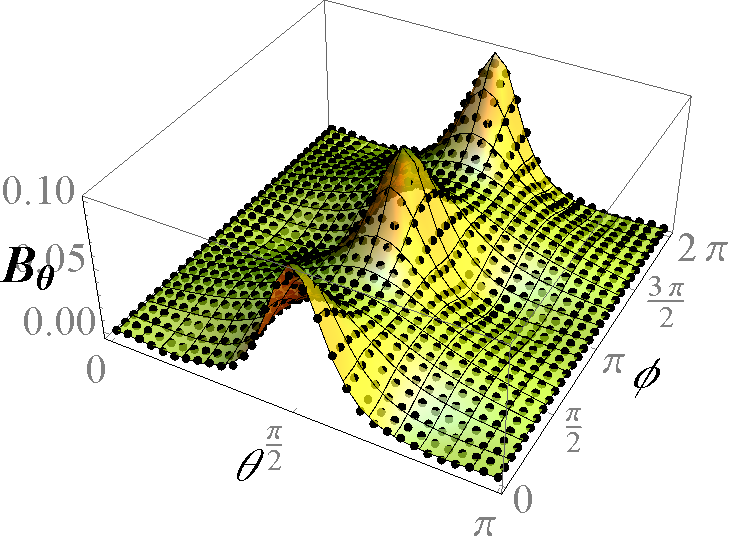}
  \caption*{(d) $r=1.4r_{max}, \frac{\tau_{num}}{\tau_{exp}} = 3.56$}
\end{minipage}
\begin{minipage}{.33\textwidth}
  \centering
  \includegraphics[width=1.0\linewidth]{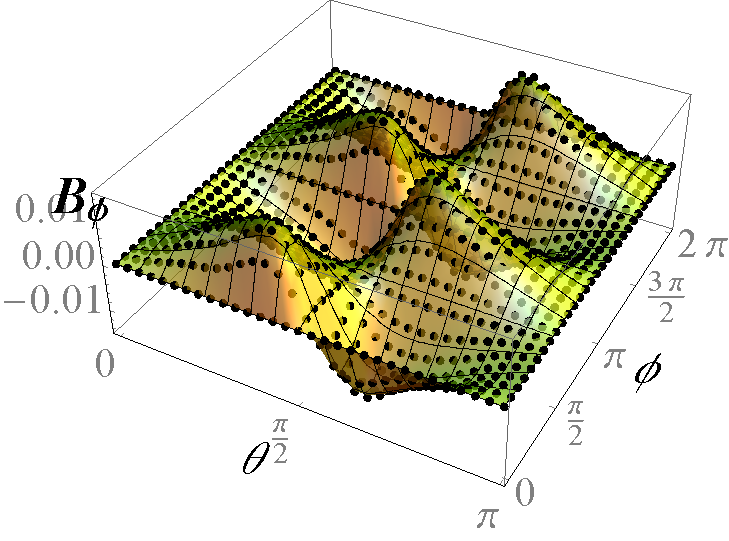}
  \caption*{(e) $r=1.4r_{max}, \frac{\tau_{num}}{\tau_{exp}} = 9.42$}
\end{minipage}

\begin{minipage}{.33\textwidth}
  \centering
  \includegraphics[width=1.0\linewidth]{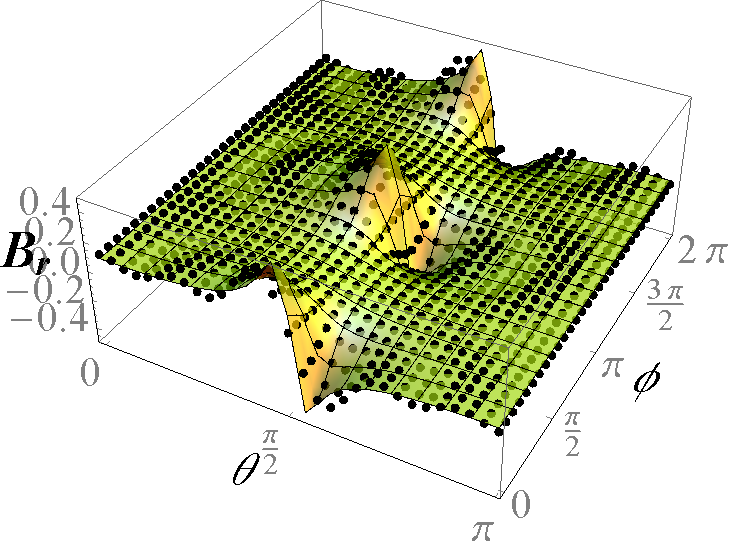}
  \caption*{(f) $r=1.1r_{max}, \frac{\tau_{num}}{\tau_{exp}} = 5.55$ } 
\end{minipage}%
\begin{minipage}{.33\textwidth}
  \centering
  \includegraphics[width=1.0\linewidth]{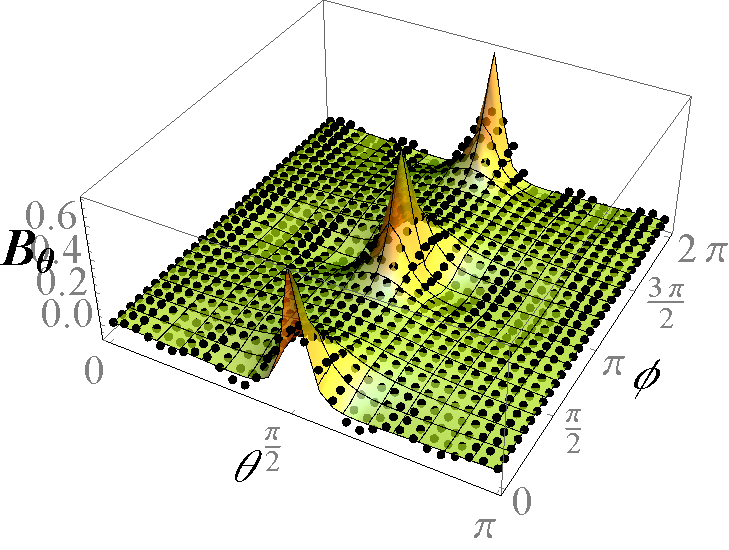}
  \caption*{(g) $r=1.1r_{max}, \frac{\tau_{num}}{\tau_{exp}} = 3.47$}
\end{minipage}
\begin{minipage}{.33\textwidth}
  \centering
  \includegraphics[width=1.0\linewidth]{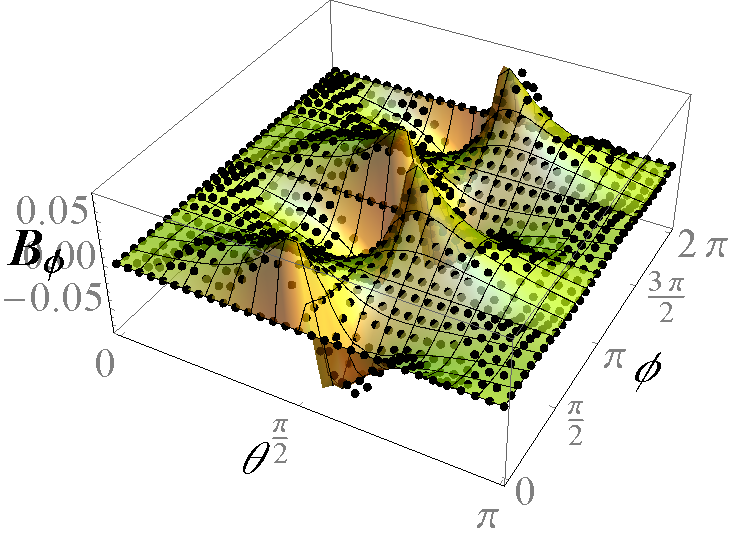}
  \caption*{(h) $r=1.1r_{max},\frac{\tau_{num}}{\tau_{exp}} = 8.91$}
\end{minipage}

\captionof{figure}{Numerical and expansion results for a harmonic deformation with $\nu=0.4$ and $p=2$. 
The surfaces represents the expanded magnetic field components evaluated at 900 points on a sphere of radius $r$. Results are contrasted with numerical integration represented by 900 black points in each plot. Results in (c), (d) and (e) corresponds to $\boldsymbol{B}$ evaluated in the sphere of radius $r=1.4r_{max}$ that is shown in (a). Plots in (f), (g) and (h) correspond to $\boldsymbol{B}$ evaluated in the sphere of radius $r=1.1r_{max}$ that is shown in (b). }
\label{Bp2nu05ComparisonFig}
\end{figure}

In general, a reduction in the time of evaluation of the magnetic field can be obtained at the cost of precision, as it is shown in Fig.~\ref{Bp2nu05ComparisonFig}-(c), (d) and (e). In these figures, the components of the magnetic field have been evaluated in a sphere of radius $r=1.4r_{max}$ (which is depicted in Fig.~\ref{Bp2nu05ComparisonFig}-(a)) by means of an expansion with $ N = 10 $ up to 6 order in $\nu$.

In this case, the strategy is accurate and the expansion represents well the numerical data with a substantial saving of computation time\footnote{In general, the performance (regarding the computational time) of the expansion is better for the $\phi$-component compared to other components since $B_{\phi}$ has fewer terms than $B_r$ and $B_{\theta}$ for a given truncation. For this term, it is not necessary to evaluate the circular loop solution because it vanishes.}. The same expansion formulas are used to evaluate the magnetic field on a sphere of radius $r=1.1r_{max}$. In this case, the expansion approach takes less time than the numerical integration but fails to correctly represent the magnetic field. An inexact description given by the truncated expansion can be observed, especially, at $ \theta = \pi/2$ when $\phi=0$ and $\phi = \pi$. This location defines the cut of the sphere with the plane z = 0, where the wire lies. At these two points, the radial and $\phi$ components of $\boldsymbol{B}$ are zero and the magnetic field $B=B_{\theta}$ presents overshoots since the point of evaluations are near to the wire. Improved accuracy can be achieved by including more terms in expansions, but this affects the computational time: as $r$ tends to $r_{max}$, the number of terms that must be included to correctly represent the field eventually cause the computational time of the expansion to exceed the numerical integration time. This is one of the limitations of the method presented in this study.



%


%

%

Besides these limitations (of the performance of the method in the vicinity of $D_{II}$), it should be noted that the expansion converges rapidly in regions of space where $r$ is large enough. Hence, it is possible to take advantage of this, as it is demonstrated in the accurate and competitive (in terms of computational cost) results in figures~\ref{accuracyFig}-(c) and \ref{Bp2nu05ComparisonFig}-(e), for $p=5$ and $p=2$, respectively. For $p=2$ the ratio between numerical and expansion computational time is $\frac{\tau_{num}}{\tau_{exp}}=8.91$. This ratio is $\frac{\tau_{num}}{\tau_{exp}}=43.64$ for $p=5$. The increase of $p$ includes more terms in the expansion (making it harder to be evaluated). Higher values of $p$ also shift the integrand into more oscillating, and therefore, the numerical integration requires more effort to converge. 

In far-from-the-wire $r/R\gg1$ regions, it is even possible to further simplify the problem since the $\phi$-coordinate dependency of the field vanishes and the standard dipole approximation works well.

\color{black}
\subsection{The $D_{II}$ region}
It is possible to propose a generalization of the equation Eq.~(\ref{expansionFormulaFinalEq}) for the region $D_{II}$. In particular, for harmonic deformations of the form $f(\phi)=\cos(p\phi)$ with $p\in\mathbb{N}$ where all roots of $r-\mathcal{R}(\phi)=0$ can be obtained from the first two roots 
\[
\beta_1(r,p) = \frac{1}{p} \arccos\left(\frac{r-R}{H}\right)\hspace{0.5cm}\mbox{and}\hspace{0.5cm}\beta_2(r,p) = -\frac{1}{p} \arccos\left(\frac{r-R}{H}\right)+\frac{2\pi}{p}.
\]
Other roots $\beta_3,\beta_4,...$ are computed by adding $2\pi/p$ multiplied by an integer depending on the root (specifically the parity of its numbering). Hence, the $j$-th root is given by
\[
\beta_j(r,p) = (\beta_2 \hspace{0.25cm}\mbox{if}\hspace{0.25cm} j \hspace{0.25cm}\in 2\mathbb{N} \hspace{0.25cm}\mbox{otherwise}\hspace{0.25cm} \beta_1 ) + (\mathfrak{n}(j)-1)\frac{2\pi}{p},
\]
where $r_{min}<r<r_{max}$ and $\mathfrak{n}(j)$ is an integer given by
\[
\mathfrak{n}(j) = \left(\frac{j}{2} \hspace{0.25cm}\mbox{if}\hspace{0.25cm} j \hspace{0.25cm}\in 2\mathbb{N} \hspace{0.25cm}\mbox{otherwise}\hspace{0.25cm} \frac{j+1}{2} \right). 
\]
The set of roots is then $\Omega(r) = \left\{ \beta_j(r,p) : j=1,...,2p  \right\}$ and it depends on $p$ and $r$. Note that $\mbox{dim}(\Omega(r))=2p=M$ is an even number. If $r$ tends to $r_{min}^{+}$ or $r_{max}^{-}$, then the roots approach each other by pairs. Thus, $\mbox{dim}(\Omega(r_{min}))=\mbox{dim}(\Omega(r_{max}))=p$ and $\mbox{dim}(\Omega(r)) = 0$ outside the $D_{II}$ region. Once the roots are located, one can write expansions formulas for the magnetic field in $D_{II}$. For instance, the $r$-component in the region $D_{II}$ takes the form\footnote{In this result it is assumed that the first two consecutive roots define angular interval $[\beta_1,\beta_2]$ belong to $\underset{r>\mathcal{R}}{\Xi}(r)$.}
\[
    B_r(\boldsymbol{r}) = \frac{\mu_o i}{4\pi} \cos\theta \sum_{n=0}^{N\rightarrow\infty} \sum_{k=0}^{\floor*{n/2}} g_{nk}^{(3)}(r,\theta) \left[\sum_{s=0}^{n} \binom{n}{s} \left(\frac{H}{R}\right)^s \sum_{m\in 2\mathbb{N}^{0}+1}^{M} \tilde{C}_{nks}^{(m)}(\phi) + \sum_{s=0}^{\infty} \binom{-\alpha-n}{s} \left(\frac{H}{R}\right)^s \sum_{m\in 2\mathbb{N}^{0}}^{M} \tilde{C}_{nks}^{(m)}(\phi)\right],
\]
where 
\[
\tilde{C}_{nks}^{(m)}(\phi) := R^2 J^{(\beta_m,\beta_{m+1})}_{q_k(n),0}[f^s] + 2 R H J^{(\beta_m,\beta_{m+1})}_{q_k(n),0}[f^{s+1}] + H^2 J^{(\beta_m,\beta_{m+1})}_{q_k(n),0}[f^{s+2}] 
\]
and $J^{(\beta,\beta')}_{m,\lambda}[F](\phi)$ are integrals depending on $\phi$, which are given by
\[
J^{(\beta,\beta')}_{m,\lambda}[F](\phi) := \int_{\beta}^{\beta'} F(\phi')\cos^m(\phi-\phi')\sin^{\lambda}(\phi-\phi')d\phi'.
\]
In general, the expansions for the magnetic are easier to evaluate in the $D_{I}$ and $D_{III}$ regions than in the $D_{II}$ region. One of the reasons for this has to do with the exact evaluation of the integrals $J^{(\beta,\beta')}_{m,\lambda}[F](\phi)$. In the $D_{I}$ and $D_{II}$ regions these integrals are $J_{m,\lambda}[F](\phi)=J^{(0,2\pi)}_{m,\lambda}[F](\phi)$ and can be calculated exactly for harmonic deformations by applying Cauchy's residue theorem in their complex plane representation (as it will discussed in the next section and formally demonstrated in Appendix \ref{JIntegralsForHarmonicDeformationAppendixSection}). This type of strategy cannot be used in the $D_{II}$ region because the $J^{(\beta,\beta')}_{m,\lambda}[F](\phi)$ cannot be represented as a closed integral in the complex plane. Additionally, the integrand is evaluated very close to the wire during the integration. This occurs, specifically, when $\phi'$ tends to one of the roots in $\Omega(r)$ and more terms in the analytical expression are required. This also implies that the evaluation through expansions becomes less practical. In the current study, the computations outside the region $D_{II}$ are mostly investigated, being the magnetic field computed via Eq.~(\ref{expansionFormulaFinalEq}).

\color{black}

\section{First-order approximation for a generic even deformation function}
\label{sec:GenericDeformation}
In this section, we shall study the first order $v=H/R$ contribution of the deformation function $f(\phi)$ to the magnetic field. This contribution is the leading term of the expansion given by Eq.~(\ref{expansionFormulaFinalEq}) since it includes the $s=0$ and $s=1$ terms involving the $J_{m,\lambda}[f]$ and $J_{m,\lambda}[\dot{f}]$ integrals. For example, the radial magnetic field of the deformed wire given by Eq.~(\ref{expansionFormulaFinalEq}) can be written as follows      
\[
    B_r(\boldsymbol{r}) = \left(B_r(\boldsymbol{r})\right)_{circle} + \frac{\mu_o i}{4\pi} \cos\theta R^2 \nu \sum_{n=0}^{N\rightarrow\infty} \sum_{k=0}^{\floor*{n/2}} g_{nk}^{(3)}(r,\theta) \left[2 + \binom{\eta(r,n)}{1}\right] J_{q_k(n),0}[f] + O(\nu^2). 
\]

Now, if the deformation function is chosen as $f=\cos(p\phi)$, then $J_{m,0}[f]$ can be computed straightforwardly from the residue theorem (see Appendix \ref{JIntegralsAppendixSection}), such that it results in
\[
J_{m,0}[\cos(p\phi)] = \beta_{m}^{(0)}(p)\cos(p\phi),   \hspace{0.5cm}\mbox{with}\hspace{0.5cm}   \beta_{m}^{(0)}(p) = \frac{2\pi}{2^m}\begin{bmatrix}
           m \\
           \frac{m-p}{2}
\end{bmatrix} .
\]
Since $J_{m,0}[\cos(p\phi)]$ is proportional to $\cos(p\phi)$, then the first order contribution of the magnetic field can be written as follows
\[
    B_r(\boldsymbol{r}) = \left(B_r(r,\theta)\right)_{circle} +  \nu \zeta_r(p,r,\theta) \cos(p\phi) + O(\nu^2), 
\]
with $\zeta_r$ a function with no dependence on $\phi$, that is given by
\[
\zeta_r(p,r,\theta) = \frac{\mu_o i}{4\pi} r R \cos\theta  \sum_{n=0}^{N\rightarrow\infty} \sum_{k=0}^{\floor*{n/2}} g_{nk}^{(3)}(r,\theta) \Lambda_r(q_k(n),p,r),
\]
with 
\[
\Lambda_r(m,p,r) = \left[2 + \binom{\eta(r,n)}{1}\right]\beta_m^{(0)}(p). 
\]

The first order deformation contributions to the other $\boldsymbol{B}(\boldsymbol{r})$ components for $f=\cos(p\phi)$ also depends on a \textit{single harmonic function}. Respectively, the angular components are given by
\[
    B_\theta(\boldsymbol{r}) = \left(B_\theta(r,\theta)\right)_{circle} +  \nu \zeta_\theta(p,r,\theta) \cos(p\phi) + O(\nu^2),
\]
and
\begin{equation}
B_\phi(\boldsymbol{r}) = \nu \zeta_\phi(p,r,\theta) \sin(p\phi) + O(\nu^2).
    \label{BphiFirstOrderHarmonicEq}
\end{equation}
This implies that components of the magnetic field up to first order in $\nu$ keep a discrete rotational symmetry of order $p$ due to the deformation function $f=\cos(p\phi)$. Since the first order deformation contribution is the most representative, then the $p$-fold symmetry can still be observed in Figs.~\ref{uPleuseursFig} and \ref{phiPleuseursFig}, even when plots in those figures include higher order terms in $\nu$. Since the deformation function $f=\cos(p\phi)$ contributes with a single harmonic to the first-order term of the magnetic field, then the result for a generic even deformation function can be generalized by using Fourier series 
\[
f(\phi) = \sum_{p=0}^\infty c_p\cos(p\phi)    \hspace{1.0cm}\mbox{even periodic function}  
\]
and the linearity of the $J_{m,\lambda}[f]$  and $J_{m,\lambda}[\dot{f}]$ integrals. 

\begin{figure}[H]
\begin{minipage}{.33\textwidth}
  \centering 
  \includegraphics[width=0.8\linewidth]{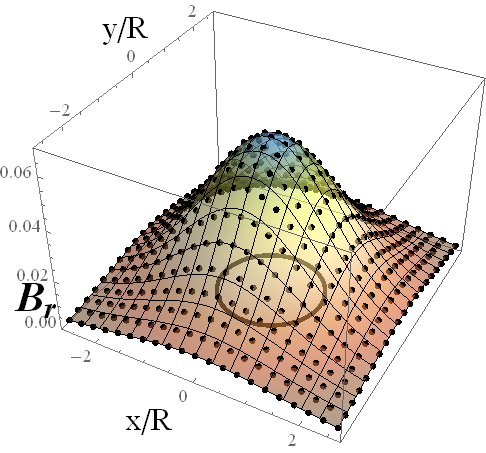}
  \caption*{(a) $\nu=0.01$, $\Xi_r = 4.3\times 10^{-9}$ } 
\end{minipage}%
\begin{minipage}{.33\textwidth}
  \centering 
  \includegraphics[width=0.8\linewidth]{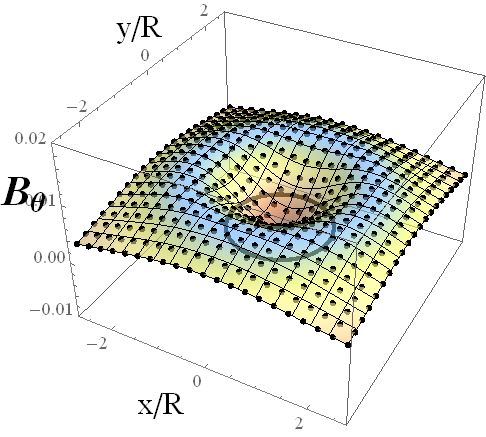}
  \caption*{(b) $\nu=0.01$, $\Xi_\theta = 9.59\times 10^{-8}$}
\end{minipage}
\begin{minipage}{.33\textwidth}
  \centering 
  \includegraphics[width=0.8\linewidth]{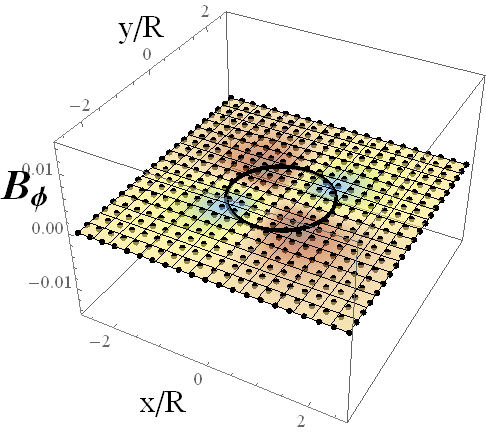}
  \caption*{(c) $\nu=0.01$, $\Xi_\phi = 1.4\times 10^{-4}$}
\end{minipage}

\begin{minipage}{.33\textwidth}
  \centering
  \includegraphics[width=0.8\linewidth]{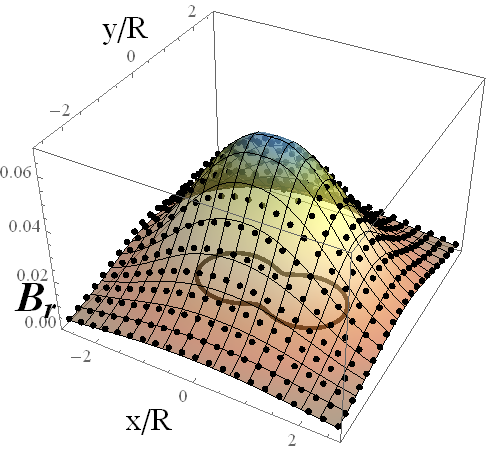}
  \caption*{(d) $\nu=0.5$, $\Xi_r = 0.0054$ } 
\end{minipage}%
\begin{minipage}{.33\textwidth}
  \centering
  \includegraphics[width=0.8\linewidth]{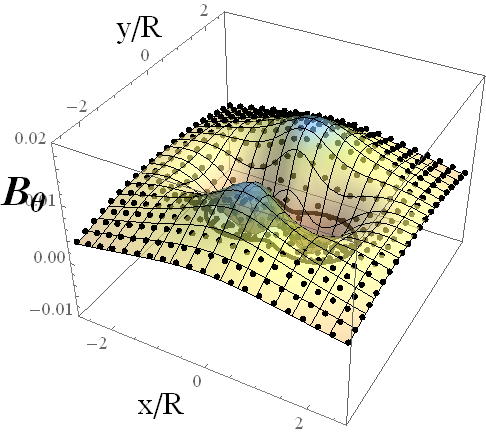}
  \caption*{(e) $\nu=0.01$, $\Xi_\theta = 0.0305$}
\end{minipage}
\begin{minipage}{.33\textwidth}
  \centering
  \includegraphics[width=0.8\linewidth]{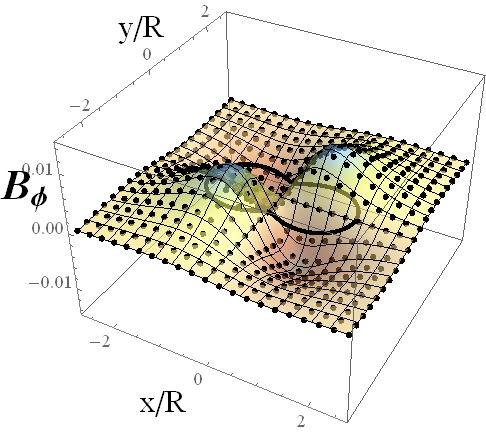}
  \caption*{(f) $\nu=0.5$, $\Xi_\phi = 0.0101$}
\end{minipage}

\begin{minipage}{.33\textwidth}
  \centering
  \includegraphics[width=0.8\linewidth]{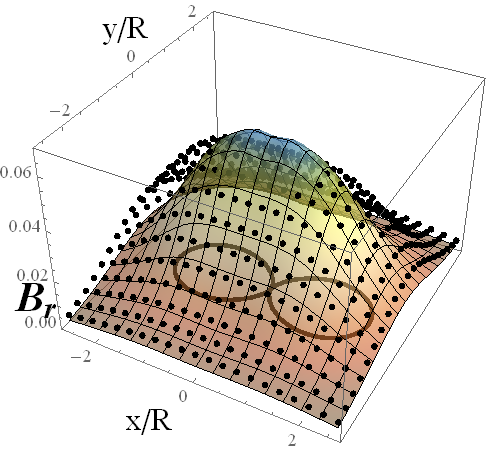}
  \caption*{(g) $\nu=0.99$, $\Xi_r = 0.0538$ } 
\end{minipage}%
\begin{minipage}{.33\textwidth}
  \centering
  \includegraphics[width=0.8\linewidth]{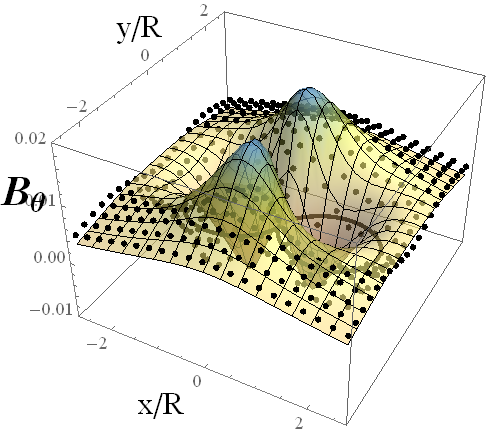}
  \caption*{(h) $\nu=0.99$, $\Xi_\theta = 0.2837$}
\end{minipage}
\begin{minipage}{.33\textwidth}
  \centering
  \includegraphics[width=0.8\linewidth]{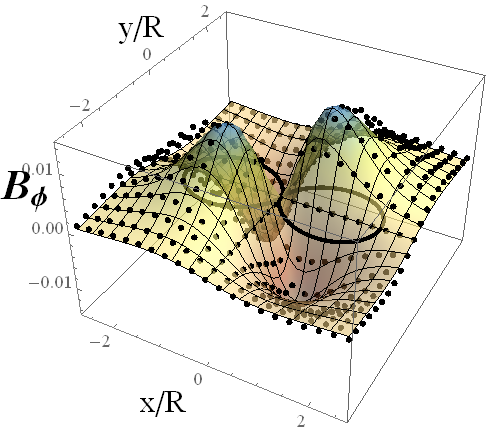}
  \caption*{(i) $\nu=0.99$, $\Xi_\phi = 0.0671$}
\end{minipage}

\captionof{figure}{First order and Numerical magnetic field due to a harmonically deformed circle with $f(\phi)=\cos(2\phi)$. Surfaces correspond to components of $\boldsymbol{B}$ evaluated a the plane $z/R=1.75$ with the first order approximation give by Eq.~(\ref{BFirstOrderInNuEq}) truncating the sum at $N=10$ and $c_p=\delta_{p,2}$. Black points correspond to numerical integration. Each plot includes the $L^2$-relative error between the numerical and first order approximation.}
\label{firstOrderp3Fig}
\end{figure}
This is,
\[
J_{m,\lambda}[f] = J_{m,\lambda}\left[\sum_{p=0}^\infty c_p\cos(p\phi)\right] = \sum_{p=0}^\infty c_p J_{m,\lambda}[\cos(p\phi)] 
\]
for the case of $J_{m,\lambda}[f]$, and a similar expression can be written for $J_{m,\lambda}[\dot{f}]$. Thus, the magnetic field due to a deformed wire by an arbitrary even function up to the first order contribution is  
\begin{equation}
\boldsymbol{B}(\boldsymbol{r}) = \left(\boldsymbol{B}(r,\theta)\right)_{circle} +  \nu \left[ \sum_{p=0}^{\infty} c_p \boldsymbol{\zeta}(p,r,\theta) \cdot \boldsymbol{\mathcal{F}}_p(\phi) \right] + O(\nu^2),
\label{BFirstOrderInNuEq}
\end{equation}
with 
\[
\boldsymbol{\mathcal{F}}_p(\phi) = \cos(p\phi) (\hat{r} + \hat{\theta}) + \sin(p\phi) \hat{\phi}
\]
and
\begin{equation}
\boldsymbol{\zeta}(p,r,\theta) = \frac{\mu_o i}{4\pi} R  \sum_{n=0}^{N\rightarrow\infty} \sum_{k=0}^{\floor*{n/2}} g_{nk}^{(3)}(r,\theta) \boldsymbol{L}_{q_k(n)}(p,r,\theta).
\label{zetaVectorFirstOrderEq}
\end{equation}
\begin{figure}[h]
\begin{minipage}{.33\textwidth}
  \centering
  \includegraphics[width=1.0\linewidth]{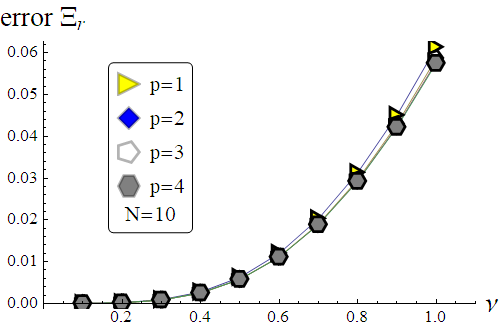}\\
  \caption*{\small(a) $\Xi_r$  } 
\end{minipage}%
\begin{minipage}{.33\textwidth}
  \centering
  \includegraphics[width=1.0\linewidth]{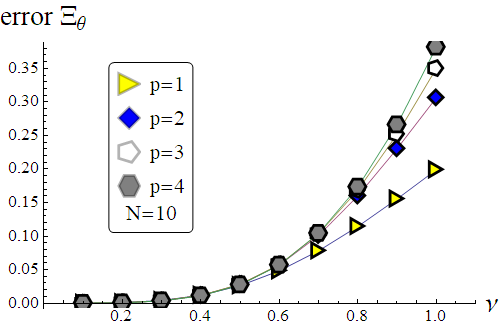}\\
  \caption*{\small(b) $\Xi_\theta$ }
\end{minipage}
\begin{minipage}{.33\textwidth}
  \centering
  \includegraphics[width=1.0\linewidth]{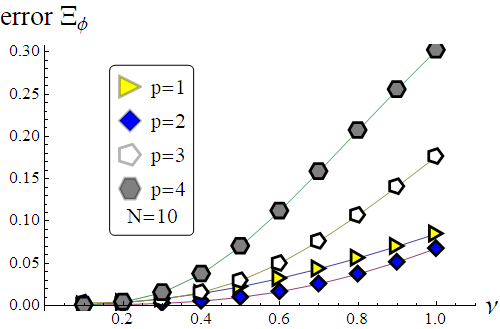}
  \caption*{\small(c) $\Xi_\phi$ }
\end{minipage}
\captionof{figure}{Error on the first order expansion of the magnetic field.}
\label{errorAFig}
\end{figure}

Here $\boldsymbol{L}_m$ is a vector with units of length whose components are
\[
(\boldsymbol{L}_m)_r = R\cos\theta\Lambda_r(m,p,r),\hspace{0.25cm}(\boldsymbol{L}_m)_\theta = r\tilde{\Lambda}_{\theta}(m,p,r)-R\sin\theta\Lambda_r(m,p,r),\hspace{0.25cm}\mbox{and}\hspace{0.25cm}(\boldsymbol{L}_m)_\phi = -r\cos\theta\Lambda_{\phi}(m,p,r).
\]

The previous terms include $\Lambda_r=Z(\beta_m^{(0)},\beta_m^{(0)},n,r)$, $\tilde{\Lambda}_\theta=Z(\beta_{m+1}^{(0)},-\beta_m^{(1)},n,r)$,  $\Lambda_{\phi}=Z(\beta_m^{(3)},\beta_{m+1}^{(4)},n,r)$, and a $Z$ function defined as follows
\[
Z(a,b,n,r) = \left[ 1+\binom{\eta(r,n)}{1} \right]a + b.
\]

It can be advisable to use a program that supports symbolic programming language to obtain analytical expressions from the analytical expansions of the magnetic field. For instance, a short code of 9 lines written in Wolfram Mathematica is shown in Appendix~\ref{CodeAppendixSection} to generate symbolic expressions of the $\phi$-component of the magnetic field for a $p$-th harmonic (this implies $c_{p'}=\delta_{p,p'}$) from Eq.~(\ref{BFirstOrderInNuEq}). Analogous codes can be written for the other components of the magnetic field. Figure~\ref{firstOrderp3Fig} shows first order and numerical fields for the harmonically deformed circle with $f(\phi)=\cos(2\phi)$. Also, points of evaluation for the error computation on the $z/R=1.75$ plane, with $x$ and $y$ ranged as plots are shown in Fig.~\ref{errorAFig}. Both figures show the $L^2$-relative error of each component of the magnetic field under the first-order approximation in $\nu$. We observe that error grows as $\nu$ is increased in all components of the magnetic field. Therefore, the first-order approximation is only valid for small values of the deformation parameter.

\color{black}
\section{Conclusions}
\label{sec:conclusions}
In this work, an efficient method to calculate the magnetic field generated by a deformed circular loop carrying a uniform electric current has been presented. 
The strategy allows to write the solution as superposition of two fields $\boldsymbol{B}(\boldsymbol{r}) = \left(\boldsymbol{B}(\boldsymbol{r})\right)_{circle} + \boldsymbol{B}^{(\nu)}(\boldsymbol{r})$, with $\left(\boldsymbol{B}(\boldsymbol{r})\right)_{circle}$ the magnetic field of the perfect circular loop that can be computed exactly and $\boldsymbol{B}^{(\nu)}(\boldsymbol{r})$ the field of the deformed counterpart.
This deformation has been represented by a periodic function $f(\phi)$ which allows us to obtain a general expression for the magnetic field in terms of the Gegenbauer polynomials.
Hence, the problem has been reduced to calculate $J_{m,\lambda}[F](\phi)$, the integrals of products including $f(\phi)$, $\dot{f}(\phi)$, and harmonic functions.

In general, the analytic expansions of the magnetic field in Eq.~(\ref{expansionFormulaFinalEq}) are useful when all the expansion coefficients are computed exactly. In other words, if $J_{m,\lambda}[f^s](\phi)$ and $J_{m,\lambda}[\dot{f}f^s](\phi)$ are solved analytically for $m, s \in \mathbb{N}$ and $\lambda=0$ or 1. In the article, we have shown the analytical integrals for pure harmonic deformations but also for anharmonic deformations in terms of Fourier series expansions of $f$.

For first-order deformations in $\nu$, the deformation only contributes to the magnetic field with a single $p$-th harmonic in the case of pure harmonic functions of the form $\cos(p\phi)$ or $\sin(p\phi)$. The zero-order term corresponding to the circular solution is axially symmetric. Thus, the first-order is the most important contribution given by the deformation. Equation~(\ref{expansionFormulaFinalEq}) explains these features of the  $p$-th-fold symmetry of the harmonic $\sin(p\phi)$. As demonstrated in Fig.~(\ref{phiPleuseursFig}) the method is not limited to first-order deformations. We generalized the result for generic even deformation functions of the first order in $\nu$.

\section*{Acknowledgments}
GT acknowledges support from Fondo de Investigaciones de la Facultad de Ciencias de la Universidad de los Andes INV-2019-84-1825 and ECOS-Nord/Minciencias C18P01. Robert Salazar thanks the support of Direcci\'on de Ciencias B\'asicas de la Universidad ECCI.

\bibliographystyle{ieeetr} 
\bibliography{bibliography.bib}

\begin{appendices}

\section{Gegenbauer Polynomials}
\label{GegenbaurePolyAppendixSection}
The summary of the main properties of the Gegenbauer polynomials $C_n :  \mathbb{C} \rightarrow \mathbb{C}$, where $n$ is the polynomial order, is presented for the convenience of the reader \cite{abramowitz1965handbook, gottlieb1992gibbs, elliott1960expansion}. 
These polynomials may be defined in terms of a generating function as 
\color{black}
\begin{equation}
\sum_{n=0}^{\infty} C_n^{(\lambda)}\left(\xi\right)\chi^n = \frac{1}{\left(1-2\xi\chi+\chi^2\right)^\lambda},
\label{genertingFunctionEq}
\end{equation}
\color{black}
where the right-hand-side of the previous expression is the generating function,  $\chi\in\mathbb{C}$, $\lambda\in\left(-1/2,\infty\right) \setminus \left\{0\right\}$, and the absolute value of $\chi$ remains $|\chi|<1$. 
The Gegenbauer Polynomials are generalizations of the Legendre Polynomials $P_n\left(\chi\right)$ and Chebyshev polynomials in the $2\left(\lambda+1\right)$-$D$ space. 
The first two polynomials are 
\[
C_{0}^{(\lambda)}(\xi) = 1,\hspace{1.0cm}\mbox{and}\hspace{1.0cm}C_{1}^{(\lambda)}(\xi) = 2\lambda\xi,
\]
respectively.
The remaining polynomials can be found with the following recurrence relation:
\[
n C_{n}^{(\lambda)}(\xi) = 2\xi (\lambda + n -1) C_{n-1}^{(\lambda)}(\xi)-2\xi (2\lambda + n -2) C_{n-2}^{(\lambda)}(\xi),
\]
where $n \geq 2$. 
These polynomials can be expressed as 
\[
C_{n}^{(\lambda)}(\xi) = \frac{(2\lambda)_n}{n!} {}_2 F_1\left(-n,2\lambda+n,\lambda+1/2;\frac{1-\xi}{2}\right),
\]
where $(x)_n$ denotes the Pochhammer symbol (rising factorial) and
\begin{equation}
{}_2 F_1(a,b,c;z) = \sum_{n=0}^\infty \frac{(a)_{n} (b)_{n}}{(c)_{n}} \frac{z^n}{n!}    
\label{hypergeometric2F1SeriesEq}
\end{equation} 
is the Gauss hypergeometric function.
This function converges when $c>0$ and $|z|<1$ or, extremely, in the unit circle $|z|=1$ if the real part $\mbox{Re}(c-a-b)>0$. 
The Gegenbauer polynomials can be also written more explicitly as follows:
\[
C_{n}^{(\lambda)}(\xi) = \sum_{k=0}^{\floor*{n/2}} (-1)^k \frac{\Gamma(n-k+\lambda)}{\Gamma(\lambda)k!(n-2k)!}\left(2\xi\right)^{n-2k} = \sum_{k=0}^{\floor*{n/2}} b_k^{\lambda} \xi^{q_k(n)}, \hspace{0.2cm}\mbox{with}\hspace{0.2cm} b_k^{(\lambda)} = (-1)^k \frac{\Gamma(n-k+\lambda)}{\Gamma(\lambda)k!(n-2k)!} 2^{q_k(n)},
\]
where $\Gamma(\cdot)$ is the gamma function, $q_k(n)=n-2k$ is the exponent, and $\floor*{n/2}$ is the floor function that takes the integer part of $n/2$. 
Additionally, the Gegenbauer polynomials are orthogonal in the $\xi \in [-1, 1]$ interval.
This is,
\begin{equation}
\int_{-1}^1 C_{n}^{(\lambda)}(\xi) C_{m}^{(\lambda)}(\xi) w^{\lambda}(\xi) d\xi = \frac{\pi 2^{1-2\lambda} \Gamma(n+2\lambda)}{n!(n+\lambda)[\Gamma(\lambda)]^2} \delta_{mn},
\label{orthogonalityConditionEq}
\end{equation}
is orthogonal with respect to the weight function  $w^{\lambda}(\xi)=(1-\xi^2)^{\lambda-1/2}$, being $\delta_{mn}$ the Kronecker delta.

The Gegenbauer polynomials are special cases of Jacobi polynomials. 
These appear naturally in the context of potential theory when a power of the inverse distance needs to be computed.
Hence, they are applied in the present work to deal with the inverse distance arising from the magnetic field calculation. 

\section{Truncated Expansion Formulas}
\label{TruncatedAppendixSection}
\subsection{Cardioid loop in the region $r_{\max} < r < \infty$}
\label{cardioidAppendixSection}
The components for $p=1$ (cardioid), taking $N=6$ for $\phi$ and $\theta$ components, and 5 for the radial component. Series are truncated up to third order in $\nu$ and $R$ is set as one.  

\begin{flushleft}
\linespread{1.5}\selectfont
\leftskip=3em  
\hspace*{-3em}$\displaystyle
B_{r}(\boldsymbol{r}) = \left(B_{r}(\boldsymbol{r})\right)_{circle} + \frac{1}{8192}\frac{\nu}{r^9}\cos (\theta ) [24 r \cos (\phi ) \sin (\theta ) (8 (24 r^4 (4+\nu
   ^2)-20 r^2 (20+27 \nu ^2)+35 (28+75 \nu ^2))+280 (4 r^2 (5+6 \nu ^2)-6 (21+50 \nu
   ^2)+3 (-25+2 r^2) \nu ^2 \cos (2 \phi )) \sin ^2(\theta )+1155 (28+75 \nu ^2) \sin ^4(\theta ))+\nu 
   (32 (64 r^6-72 r^4 (8+\nu ^2)+180 r^2 (10+3 \nu ^2)-35 (112+45 \nu ^2))+240 (3
   (2352+945 \nu ^2+8 r^4 (8+\nu ^2)-56 r^2 (10+3 \nu ^2))+2 (8 r^4 (6+\nu ^2)-84 r^2 (5+2
   \nu ^2)+63 (28+15 \nu ^2)) \cos (2 \phi )) \sin ^2(\theta )+630 (16 (6 r^2 (5+2 \nu ^2)-11
   (28+15 \nu ^2)) \cos (2 \phi )+3 (24 r^2 (10+3 \nu ^2)-22 (112+45 \nu ^2)+(-55+4 r^2)
   \nu ^2 \cos (4 \phi ))) \sin ^4(\theta )+346500 r \nu  \cos (3 \phi ) \sin ^5(\theta )+15015 (224+90 \nu ^2+6 (28+15
   \nu ^2) \cos (2 \phi )+9 \nu ^2 \cos (4 \phi )) \sin ^6(\theta ))] + O(\nu^4).
$
\end{flushleft}

\begin{flushleft}
\linespread{1.5}\selectfont
\leftskip=3em
\hspace*{-3em}$\displaystyle
B_{\theta}(\boldsymbol{r}) = \left(B_{\theta}(\boldsymbol{r})\right)_{circle} + \tan\theta B_{r}^{(\nu)}(\boldsymbol{r}) + \frac{3}{32768} \nu  \frac{1}{r^8} [\nu  \sin (\theta ) (8 r (9975-960 r^2+512 r^4-420 (-15+16 r^2) \cos (2
   \theta )+17325 \cos (4 \theta ))-80 r (-385-144 r^2+84 (5+4 r^2) \cos (2 \theta )-1155 \cos (4 \theta )) \cos (2
   \phi )+105 \nu  (-12 (33+8 r^2) \cos (2 \theta )+11 (-3+32 r^2+39 \cos (4 \theta ))) \cos (3 \phi ) \sin
   (\theta ))+8 \cos (\phi ) (-32 (8 r^4 (4+\nu ^2)-20 r^2 (4+3 \nu ^2)+35 (4+5 \nu
   ^2))+160 (8 r^4 (4+\nu ^2)+315 (2+3 \nu ^2)-49 r^2 (4+3 \nu ^2)-42 r^2 \nu ^2 \cos (2 \phi
   )) \sin ^2(\theta )+420 (12 r^2 (8+7 \nu ^2)-11 (68+105 \nu ^2)) \sin ^4(\theta )+15015 (16+25
   \nu ^2) \sin ^6(\theta ))] + O(\nu^4).
$
\end{flushleft}

\begin{flushleft}
\linespread{1.5}\selectfont
\leftskip=3em 
\hspace*{-3em}$\displaystyle
B_{\phi}(\boldsymbol{r}) =  \frac{3 \nu}{4096}   \frac{1}{r^7} \cos (\theta ) \sin (\phi ) [-320 r+1024 r^3+915 r \nu ^2+256 r^3 \nu ^2-140 r (16+27 \nu^2) \cos (2 \theta )  + 945 r \nu ^2 \cos (4 \theta )+315 r \nu ^2 \cos (4 \theta -2 \phi )-1260 r \nu ^2 \cos (2 (\theta -\phi ))+1890 r
   \nu ^2 \cos (2 \phi )   -1260 r \nu ^2 \cos (2 (\theta +\phi ))+315 r \nu ^2 \cos (2 (2 \theta +\phi ))+1400 \nu  \sin (\theta -\phi )+1920 r^2
   \nu  \sin (\theta -\phi )  -4200 \nu  \sin (3 \theta -\phi )+1400 \nu  \sin (\theta +\phi )+1920 r^2 \nu  \sin (\theta +\phi )-4200 \nu  \sin
   (3 \theta +\phi )]
 + O(\nu^4).
$
\end{flushleft}

\subsection{Harmonically deformed loop with $p=3$ in the region $0 \leq r < r_{\min}$}
\label{p3appendixSection}

\begin{flushleft}
\linespread{1.5}\selectfont
\leftskip=3em  
\hspace*{-3em}$\displaystyle
B_{r}^{(\nu)}(\boldsymbol{r}) = \frac{1}{4096} \nu  \cos (\theta ) (-16 \nu  (-360 r^4 (5+7 \nu ^2)-32 (2+9 \nu ^2)+72 r^2 (8+15 \nu
   ^2)+35 r^6 (112+135 \nu ^2))+360 r^2 \nu  (-112 r^2 (5+7 \nu ^2)+8 (8+15 \nu ^2)+21 r^4(112+135 \nu ^2)) \sin ^2(\theta )-4480 r^3 (4-27 \nu ^2+27 r^2 (-1+8 \nu ^2)) \cos (3 \phi ) \sin
   ^3(\theta )-1890 r^4 \nu  (-24 (5+7 \nu ^2)+11 r^2 (112+135 \nu ^2)) \sin ^4(\theta )+166320 r^5 (-1+8
   \nu ^2) \cos (3 \phi ) \sin ^5(\theta )+3003 r^6 \nu  (560+675 \nu ^2+(28+45 \nu ^2) \cos (6 \phi )) \sin
   ^6(\theta )) + O(\nu^4).$
\end{flushleft}

\begin{flushleft}
\linespread{1.5}\selectfont
\leftskip=3em 
\hspace*{-3em}$\displaystyle
B_{\theta}^{(\nu)}(\boldsymbol{r}) =  \tan\theta B_{r}^{(\nu)}(\boldsymbol{r}) + \frac{3}{4096} \nu  r^2 \sin (\theta ) (35 r \cos (3 \phi ) \sin (\theta ) (-16 (8-36 r^2+99 r^4) - 6 r^2 (156-168 \nu^2 + 11 r^2 (-98+135 \nu ^2)) \sin ^2(\theta )  +429 r^4 (-13+27 \nu ^2) \sin ^4(\theta ))+8 \nu  (32 (12-75 r^2+245 r^4) -840 r^2 (-5+42 r^2) \sin ^2(\theta )+231 r^4 (140+13 \cos (6 \phi )) \sin ^4(\theta ))) + O(\nu^4).
$
\end{flushleft}

\begin{flushleft}
\linespread{1.5}\selectfont
\leftskip=3em 
\hspace*{-3em}$\displaystyle
B_{\phi}(\boldsymbol{r}) =  \frac{3\nu}{4096}   \frac{\sin (\theta )}{r^8} (r \nu  (9975-960 r^2+512 r^4-420 (-15+16 r^2) \cos (2 \theta ) +17325 \cos(4 \theta )) +35 \cos (3 \phi ) \sin (\theta ) (64 (9-2 r^2)  +6 (4 r^2 (16+9 \nu ^2)-11 (52+45 \nu^2)) \sin ^2(\theta )+429 (8+9 \nu ^2) \sin ^4(\theta )))
 + O(\nu^4).
$
\end{flushleft}

\section{J's integrals with harmonic deformation $f=\cos(p\phi)$}
\label{JIntegralsAppendixSection}
Starting from Eq.~(\ref{JLambdaZeroEq}) we obtain for $\lambda=0$
\[
J_{m,0}[f](\phi) = \frac{2\pi}{2^{s+m}} \sum_{k=0}^s\binom{1}{k} \begin{bmatrix}
           m \\
           l'_{k 1 m p 0 0}
         \end{bmatrix} \cos[(2k-1)p\phi] = \frac{\pi}{2^m}  \left\{
         \begin{bmatrix}
           m \\
           \frac{m-p}{2}
        \end{bmatrix}   +  \begin{bmatrix}
           m \\
           \frac{m+p}{2}
\end{bmatrix}\right\} \cos(p\phi)
\]
which is zero when $m<p$ or $m \pm p$ is an odd number according to the definition given by Eq.~(\ref{conditionatedbinomEq}). The result is not zero when $m-p$ is a positive even number say $m-p=2t \in 2\mathbb{N}$, thus
\[
\begin{bmatrix}
           m \\
           \frac{m-p}{2}
\end{bmatrix} = \begin{bmatrix}
           m \\
           t
        \end{bmatrix} = \binom{m}{t}.
\]
On the other hand
\[
\begin{bmatrix}
           m \\
           \frac{m+p}{2}
\end{bmatrix} = \begin{bmatrix}
           m \\
           m-t
\end{bmatrix} = \frac{m!}{(m-t)!t!} = \binom{m}{t} = \begin{bmatrix}
           m \\
           \frac{m-p}{2}
\end{bmatrix}
\]
hence
\[
J_{m,0}[f](\phi) =  \beta_m^{(0)}(p)  \cos(p\phi) \hspace{0.5cm}\mbox{with}\hspace{0.5cm} \beta_m^{(0)}(p) =  \frac{2\pi}{2^m}  
         \begin{bmatrix}
           m \\
           \frac{m-p}{2}
        \end{bmatrix}.
\]
If $\lambda=1$ only four terms of Eq.~(\ref{JLambdaOneEq}) contribute
\[
J_{m,1}[f](\phi) = \frac{2\pi}{2^{m+2}}  \left\{
         \begin{bmatrix}
           m \\
           \frac{-p+m+1}{2}
        \end{bmatrix}   -  \begin{bmatrix}
           m \\
           \frac{-p+m-1}{2}
\end{bmatrix}-
\begin{bmatrix}
           m \\
           \frac{p+m+1}{2}
\end{bmatrix}+\begin{bmatrix}
           m \\
           \frac{p+m-1}{2}
\end{bmatrix}
\right\} \sin(p\phi) .
\]
Now 
\[
\begin{bmatrix}
           m \\
           \frac{-p+m+1}{2}
\end{bmatrix} = \begin{bmatrix}
           m \\
           \frac{p+m-1}{2}
        \end{bmatrix} 
        \hspace{0.5cm}\mbox{,}\hspace{0.5cm}\begin{bmatrix}
           m \\
           \frac{-p+m-1}{2}
        \end{bmatrix} = \begin{bmatrix}
           m \\
           \frac{p+m+1}{2}
        \end{bmatrix}   
\]
then $J_{m,1}[f](\phi)$ can be simplified as follows
\[
J_{m,1}[f](\phi) = \frac{4\pi}{2^{m+2}}  \left\{
\begin{bmatrix}
           m \\
           \frac{p+m-1}{2}
\end{bmatrix}-
\begin{bmatrix}
           m \\
           \frac{p+m+1}{2}
\end{bmatrix}
\right\} \sin(p\phi) 
\]
The conditioned binomials are not zero when $p+m=2t+1$ is a positive odd number
\[
\begin{bmatrix}
           m \\
           \frac{p+m-1}{2} 
\end{bmatrix} 
= \binom{m}{t}\hspace{0.5cm}\mbox{and}\hspace{0.5cm} \begin{bmatrix}
           m \\
           \frac{p+m+1}{2} 
\end{bmatrix} 
= \binom{m}{t+1} = \frac{m-t}{t+1} \binom{m}{t} = \frac{m-p+1}{p+m+1} \begin{bmatrix}
           m \\
           \frac{p+m-1}{2} 
\end{bmatrix}
\]
therefore
\[
J_{m,1}[f](\phi) = \beta_{m}^{(3)} \sin(p\phi)  \hspace{0.5cm}\mbox{with}\hspace{0.5cm}  \beta_{m}^{(3)} = \frac{\pi}{2^m}\frac{2p}{p+m+1}  
\begin{bmatrix}
           m \\
           \frac{p+m-1}{2}  
\end{bmatrix}  .
\]
The integrals $J_{m,0}[\dot{f}]$ and $J_{m,1}[\dot{f}]$ can be computed similarly from Eqs.~(\ref{JDotLambdaZeroEq}) and (\ref{JDotLambdaOneEq}), the result is
\[
J_{m,1}[\dot{f}] = \beta_{m}^{(1)} \cos(p\phi)   \hspace{1.0cm}\mbox{and}\hspace{1.0cm} J_{m,0}[\dot{f}] = \beta_{m}^{(4)} \sin(p\phi)
\]
with 
\[
\beta_{m}^{(1)} = \frac{2\pi}{2^m}\frac{p^2}{p+m+1}\begin{bmatrix}
           m \\
           \frac{p+m-1}{2}
\end{bmatrix} \hspace{0.5cm}\mbox{and}\hspace{0.5cm} \beta_{m}^{(4)} = -\frac{4\pi p}{2^{m+1}}\begin{bmatrix}
           m \\
           \frac{p+m}{2}
\end{bmatrix} .
\]

\section{Short code to compute the first order term of $B_\phi$}
\label{CodeAppendixSection}
\begin{figure}[h]
\begin{minipage}{1.0\textwidth}
  \centering 
  \includegraphics[width=0.99\linewidth]{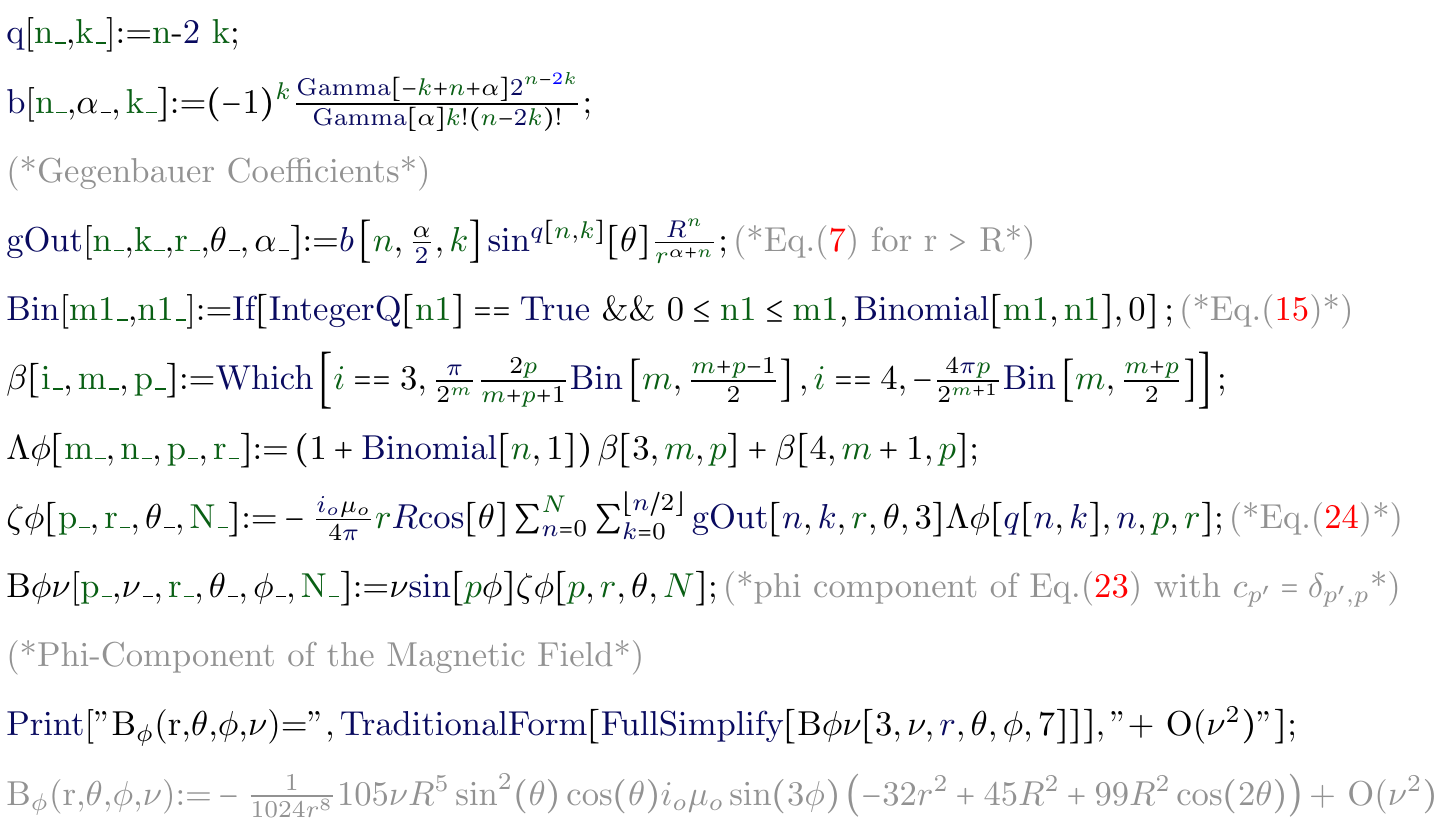}
\end{minipage}
\captionof{figure}{$\mathtt{Mathematica}$ code to compute the first order term of $B_\phi$.}
\label{listingFig}
\end{figure}

A short \texttt{Mathematica} code that symbolically computes the $\phi$-component of the magnetic field is demonstrated in Fig~\ref{listingFig}. This code is restrained to the first order truncation in $\nu$ and the $f(\phi)=\cos(p\phi)$ deformation function. In the code, $B_\phi(\boldsymbol{r})$ is given by the $B\phi\nu[p,\nu,r,\theta,\phi]$ function and the output in the last cell is computed by defining $N=7$ and $p=3$. This code can be further modified to include the remaining components of the magnetic field.

\end{appendices}






\end{document}